\RequirePackage{fix-cm}
\documentclass[twocolumn,epjc3]{svjour3}  
\smartqed  
\RequirePackage{graphicx}
\RequirePackage{subcaption}
\usepackage{amsmath}
\usepackage{amssymb}
\usepackage{multirow}
\usepackage{lineno} 
\usepackage{siunitx}
\newcommand{\units}[1]{\,\mathrm{#1}}
\journalname{Eur. Phys. J. C}
\begin{document}

\title{Validating the performance of the Radio Neutrino Observatory in Greenland using cosmic-ray air showers
}

\titlerunning{Air showers signals in RNO-G}        

\author{
S.~Agarwal\thanksref{KU} \and
J.~A.~Aguilar\thanksref{ULB} \and
N.~Alden\thanksref{UC} \and
S.~Ali\thanksref{KU} \and
P.~Allison\thanksref{OSU} \and
M.~Betts\thanksref{PSUPhysics,PSUAstro} \and
D.~Besson\thanksref{KU} \and
A.~Bishop\thanksref{WIPAC} \and
O.~Botner\thanksref{UppPh} \and
S.~Bouma\thanksref{ECAP} \and
S.~Buitink\thanksref{VUBastro,RU} \and
R.~Camphyn\thanksref{ULB} \and
J.~Chan\thanksref{WIPAC} \and
S.~Chiche\thanksref{ULB} \and
B.~A.~Clark\thanksref{UMD} \and
A.~Coleman\thanksref{UppPh} \and
K.~Couberly\thanksref{KU} \and
S.~de Kockere\thanksref{VUBelem} \and
K.~D.~de Vries\thanksref{VUBelem} \and
C.~Deaconu\thanksref{UC} \and
P.~Giri\thanksref{UNL} \and
C.~Glaser\thanksref{UppPh,TUDo} \and
T.~Gl{\"u}senkamp\thanksref{UppPh} \and
H.~Gui\thanksref{OSU} \and
A.~Hallgren\thanksref{UppPh} \and
S.~Hallmann\thanksref{DESY,ECAP} \and
J.~C.~Hanson\thanksref{Whittier} \and
K.~Helbing\thanksref{BUW} \and
B.~Hendricks\thanksref{PSUPhysics,PSUIGC} \and
J.~Henrichs\thanksref{DESY,ECAP} \and
N.~Heyer\thanksref{UppPh} \and
C.~Hornhuber\thanksref{KU} \and
E.~Huesca Santiago\thanksref{DESY} \and
K.~Hughes\thanksref{OSU} \and
A.~Jaitly\thanksref{DESY,ECAP} \and
T.~Karg\thanksref{DESY} \and
A.~Karle\thanksref{WIPAC} \and
J.~L.~Kelley\thanksref{WIPAC} \and
C.~Kopper\thanksref{ECAP} \and
M.~Korntheuer\thanksref{ULB,VUBelem} \and
M.~Kowalski\thanksref{DESY,HU} \and
I.~Kravchenko\thanksref{UNL} \and
R.~Krebs\thanksref{PSUPhysics,PSUIGC} \and
M.~Kugelmeier\thanksref{WIPAC} \and
D.~Kullgren\thanksref{PSUPhysics,PSUIGC} \and
R.~Lahmann\thanksref{ECAP} \and
C.-H. Liu\thanksref{UNL} \and
Y.~Liu\thanksref{OSU} \and
M.~J.~Marsee\thanksref{UAlabama} \and
K.~Mulrey\thanksref{RU} \and
M.~Muzio\thanksref{WIPAC} \and
A.~Nelles\thanksref{DESY,ECAP} \and
A.~Novikov\thanksref{UD} \and
A.~Nozdrina\thanksref{OSU} \and
E.~Oberla\thanksref{UC} \and
B.~Oeyen\thanksref{Ghent} \and
N.~Punsuebsay\thanksref{UD} \and
L.~Pyras\thanksref{DESY,Utah} \and
M.~Ravn\thanksref{UppPh} \and
A.~Rifaie\thanksref{BUW} \and
D.~Ryckbosch\thanksref{Ghent} \and
F.~Schl{\"u}ter\thanksref{ULB} \and
O.~Scholten\thanksref{VUBelem,UG} \and
D.~Seckel\thanksref{UD} \and
M.~F.~H.~Seikh\thanksref{KU} \and
Z.~S.~Selcuk\thanksref{DESY,ECAP} \and
J.~Stachurska\thanksref{Ghent} \and
J.~Stoffels\thanksref{VUBelem} \and
S.~Toscano\thanksref{ULB} \and
D.~Tosi\thanksref{WIPAC} \and
J.~Tutt\thanksref{PSUAstro} \and
D.~J.~Van Den Broeck\thanksref{VUBelem,VUBastro} \and
N.~van Eijndhoven\thanksref{VUBelem} \and
A.~G.~Vieregg\thanksref{UC} \and
A.~Vijai\thanksref{UMD} \and
D.~Washington\thanksref{PSUPhysics,PSUAstro} \and
C.~Welling\thanksref{PSUPhysics,PSUAstro,PSUIGC} \and
D.~R.~Williams\thanksref{UAlabama} \and
P.~Windischhofer\thanksref{UC} \and
S.~Wissel\thanksref{PSUPhysics,PSUAstro,PSUIGC} \and
R.~Young\thanksref{KU} \and
A.~Zink\thanksref{ECAP}
}

\thankstext{e1}{e-mail: jakob.henrichs@desy.de, authors@rno-g.org}

\authorrunning{RNO-G Collaboration} 

\institute{
 University of Kansas, Dept.\ of Physics and Astronomy, Lawrence, KS 66045, USA \label{KU} \and
 Universit\'e Libre de Bruxelles, Science Faculty CP230, B-1050 Brussels, Belgium \label{ULB} \and
 Dept.\ of Physics, Dept.\ of Astronomy \& Astrophysics, Enrico Fermi Inst., Kavli Inst.\ for Cosmological Physics, University of Chicago, Chicago, IL 60637, USA \label{UC} \and
 Dept.\ of Physics, Center for Cosmology and AstroParticle Physics, Ohio State University, Columbus, OH 43210, USA \label{OSU} \and
 Dept.\ of Physics, Pennsylvania State University, University Park, PA 16802, USA \label{PSUPhysics} \and
 Dept.\ of Astronomy and Astrophysics, Pennsylvania State University, University Park, PA 16802, USA \label{PSUAstro} \and
 Wisconsin IceCube Particle Astrophysics Center (WIPAC) and Dept.\ of Physics, University of Wisconsin-Madison, Madison, WI 53703,  USA \label{WIPAC} \and
 Uppsala University, Dept.\ of Physics and Astronomy, Uppsala, SE-752 37, Sweden \label{UppPh} \and
 Erlangen Centre for Astroparticle Physics (ECAP), Friedrich-Alexander-University Erlangen-N\"urnberg, 91058 Erlangen, Germany \label{ECAP} \and
 Vrije Universiteit Brussel, Astrophysical Institute, Pleinlaan 2, 1050 Brussels, Belgium \label{VUBastro} \and
 Dept.\ of Astrophysics/IMAPP, Radboud University, PO Box 9010, 6500 GL, The Netherlands \label{RU} \and
 Dept.\ of Physics, University of Maryland, College Park, MD 20742, USA \label{UMD} \and
 Vrije Universiteit Brussel, Dienst ELEM, B-1050 Brussels, Belgium \label{VUBelem} \and
 Dept.\ of Physics and Astronomy, Univ.\ of Nebraska-Lincoln, NE, 68588, USA \label{UNL} \and
 Dept.\ of Physics, TU Dortmund University, Dortmund, Germany \label{TUDo} \and
 Deutsches Elektronen-Synchrotron DESY, Platanenallee 6, 15738 Zeuthen, Germany \label{DESY} \and
 Whittier College, Whittier, CA 90602, USA \label{Whittier} \and
 Dept. of Physics, University of Wuppertal  D-42119 Wuppertal, Germany \label{BUW} \and
 Center for Multimessenger Astrophysics, Inst.\ of Gravitation and the Cosmos, Pennsylvania State University, University Park, PA 16802, USA \label{PSUIGC} \and
 Institut f\"ur Physik, Humboldt-Universit\"at zu Berlin, 12489 Berlin, Germany \label{HU} \and
 Dept.\ of Physics and Astronomy, University of Alabama, Tuscaloosa, AL 35487, USA \label{UAlabama} \and
 Dept.\ of Physics and Astronomy, University of Delaware, Newark, DE 19716, USA \label{UD} \and
 Ghent University, Dept.\ of Physics and Astronomy, B-9000 Gent, Belgium \label{Ghent} \and
 Dept.\ of Physics and Astronomy, University of Utah, Salt Lake City, UT 84112, USA \label{Utah} \and
 Kapteyn Institute, University of Groningen, PO Box 800, 9700 AV, The Netherlands \label{UG}
}

\date{Received: date / Accepted: date}

\maketitle

\begin{abstract}
The Radio Neutrino Observatory in Greenland (RNO-G) is currently under construction with the aim to detect neutrinos with energies beyond $\sim 10 \units{PeV}$. A critical part of early detector commissioning is the study of detector characteristics and potential backgrounds, for which cosmic rays play a crucial role. In this article, we report that the number of cosmic rays detected with RNO-G's shallow antennas is consistent with expectations. We further verified the agreement in the observed cosmic-ray signal shape with expectations from simulations after careful treatment of the detector systematics. Finally, we find that the reconstructed arrival direction, energy, and polarization of the cosmic-ray candidates agrees with expectations. Throughout this study, we identified detector shortcomings that are mitigated going forward. Overall, the analysis presented here is an essential first step towards validating the detector and high-fidelity neutrino detection with RNO-G in the future. 
\end{abstract}

\section{Introduction}
\label{intro}
Other than the exceptional ultra-high-energy neutrino (UHEN, $E_\nu> 10^{16} \units{eV}$) reported by the KM3NeT collaboration with an estimated neutrino energy of $220\units{PeV}$ \cite{KM3NeT:2025npi}, no other observation of the neutrino flux above $10^{16}$~eV has been reported. 
However, at energies below $10^{16} \units{eV}$, neutrino astronomy is a well established field. In recent years, the IceCube collaboration has identified astrophysical neutrino sources, such as a blazar \cite{icecube_txs} and a nearby active galaxy \cite{icecube_ngc}, detected neutrinos emitted from our Galaxy \cite{icecube_galaxy}, and measured the diffuse neutrino flux up to an energy of $10 \units{PeV}$ \cite{icecube_diffuse}. In the ultra-high-energy regime the most stringent limits on the neutrino flux come from IceCube and the Pierre Auger Observatory \cite{IceCubeCollaborationSS:2025jbi,PierreAuger:2019ens} and are in mild tension with the observation by KM3NeT. 

Due to the low flux of UHEN, it is necessary to construct detectors with even larger detection volumes than KM3NeT \cite{km3net_letter_of_intent} or IceCube \cite{Aartsen_2017} to detect them. A cost-effective approach is to deploy radio antennas in naturally occurring ice, thereby exploiting the radio-transparent properties of the medium as the detection volume. When an UHEN interacts with the ice, a particle cascade is created that produces electromagnetic emission in the MHz range via the so-called Askaryan effect \cite{Askaryan}. Thanks to the large attenuation length of e.g.\ the Greenlandic ice of $\mathcal{O}(1\units{km})$ \cite{rnog_att_len}, the produced radio emission can propagate over large distances before being detected with radio antennas. This detection principle enables the observation of a large volume of ice with only a few antennas.

Previous in-ice radio experiments, ARA and ARIANNA, have successfully demonstrated the radio detection technique in the ultra-high energy regime \cite{ARA_flux,ARIANNA}. In addition, balloon-borne experiments equipped with radio antennas to observe large areas of ice have provided constraints on the UHEN flux \cite{anita_flux}.

An important background for in-ice UHEN radio detectors is the radio emission from ultra-high-energy cosmic-ray (UHECR, $E_\mathrm{CR}> 10^{16} \units{eV}$) air showers. When an UHECR interacts in the atmosphere, a in-air particle cascade (air shower) is created, which produces radio emission, predominately via the geomagnetic mechanism \cite{geomagnetic}. The resulting radio emission can refract into the ice and propagate to the radio antennas, becoming a background to the observation of an UHEN signal \cite{ARIANNA_air_shower}. In addition, it was demonstrated that incomplete air-showers further developing after entering the ice create an additional source of background related to cosmic-ray air showers \cite{deVries:2015oda,Coleman:2024cln,DeKockere:2022bto}. Finally, muons from the air shower can penetrate the ice and stochastically lose energy, producing a particle cascade effectively indistinguishable from an UHEN \cite{Pyras:2023crm}. Therefore, it is crucial to understand signals from air showers and, whenever possible, to veto them. On the other hand, signals from UHECR air showers have similar characteristics to UHEN signals, as both produce electric fields on very short (nanosecond) time scales. Consequently, these signals provide an opportunity to test the detector under realistic conditions, effectively serving as a unique ``test beam''. Moreover, radio UHEN detectors with a large effective volume also have a large effective area for the detection of UHECR, which makes it possible to measure and study their properties.

\begin{figure}
  \includegraphics[width=0.5\textwidth]{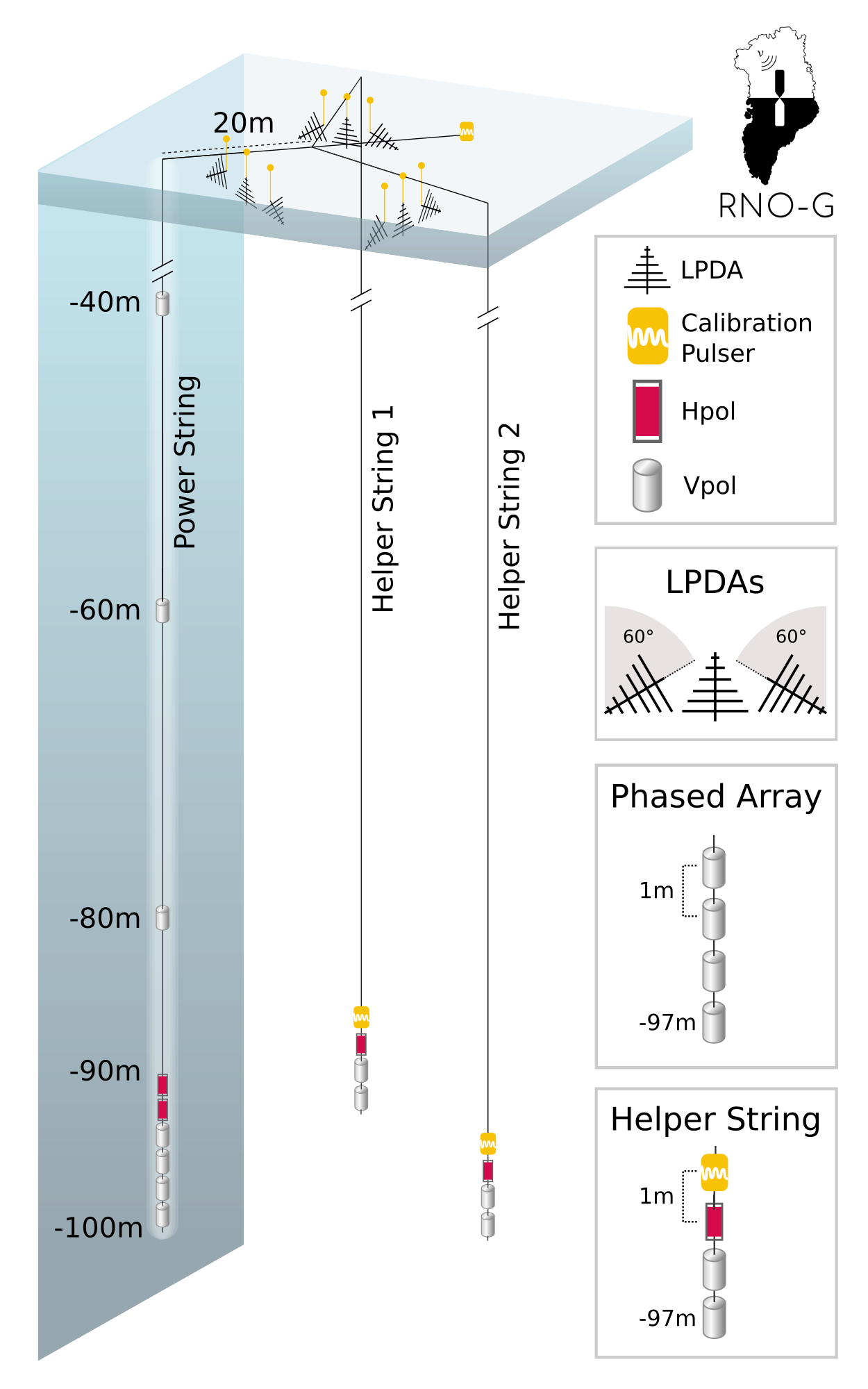}
\caption{The figure shows a schematic of the first 7 RNO-G stations. 24 antennas of three types (LPDA, Hpol, and Vpol) are distributed in the shallow ice and in three boreholes. }
\label{fig:rnog_station}
\end{figure}

Building on experience from existing or past in-ice radio neutrino experiments \cite{RICE,ARA,ARIANNA}, the Radio Neutrino Observatory in Greenland (RNO-G) is currently being constructed and will be sensitive to UHEN with energies above $10 \units{PeV}$ \cite{rnog_whitepaper}. RNO-G is located on top of the Greenlandic ice sheet at a height of approx. $3200\units{m}$, close to Summit Station. Once completed, the detector will consist of 35 independently operating stations, leading to an effective volume of $\sim100 \units{km^3}$, covering an area of $\sim50\units{km^2}$ and thus will be the largest in-ice radio UHEN detector \cite{rnog_whitepaper}. Currently, 8 stations (labeled with a two-digit number) are deployed and operational. Seven of these stations have the same geometry as shown in Fig.~\ref{fig:rnog_station} \cite{rnog_instrument_paper}. For the newest station, the antennas closest to the surface have been rearranged to better measure coincident signals in all antennas, which will be the design going forward for future stations. 

For this analysis, only the first 7 stations with the same geometry will be considered. All stations are entirely powered by renewable energy with a limited power budget, which drives the design choices made for electronic components. Currently, the power primarily comes from solar panels, limiting the data-taking period to the Arctic summer months. However, for testing purposes some stations are equipped with a wind turbine to extend the data-taking period further into the winter months. A summary of the detector performance can be found in \cite{rnog_instrument_paper}.

Each station consists of a deep and shallow part. For the deep part, custom-built antennas are deployed in three boreholes down to a depth of $100 \units{m}$. Additionally, two calibration pulsers are co-deployed with the antennas. The shallow antennas are commercially available log-periodic-dipole antennas (LPDA), installed in hand-dug trenches near the surface. In total, 9 LPDAs make up the shallow part of the detector, with 6 of them facing downwards. These downward-facing antennas point with their most sensitive part into the ice and hence contribute to the effective volume sensitive to neutrinos. The other three LPDAs are facing upwards and are used to detect and veto backgrounds, like UHECR air showers. For this reason, the three upward-facing antennas are the most important ones for this UHECR analysis and are referred to as channels 13, 16, and 19. The newest station has a different shallow geometry, where only 8 LPDAs are deployed with 4 oriented upward and 4 downward, with an additional dipole antenna at shallow depth. 

Although the electric field from UHECR air-showers is intrinsically very short in time and has a bipolar structure \cite{Huege:2013vt}, it will be dispersed by the antenna, amplification and digitization responses. As a result, the expected signal is a pulse with an oscillating signature, with the high frequencies appearing at the beginning and the low frequencies at the end (for an example, see Fig.~\ref{fig:templates}).
Due to the high elevation of Summit Station, the detector is located at a vertical depth of $\sim 700 \units{g/cm^2}$, which is close to the shower maximum, $\mathrm{X_{max}}$, of air showers with energies of $\sim 10^{17}\units{eV}$ \cite{shower_xmax}. As a result, RNO-G will also be able to observe signatures of air-shower cores that impact the ice surface and continue to develop a particle cascade in the ice that produces Askaryan emission, as reported in \cite{ara_impacting_shower}.

In this work we will present a full analysis to identify signals from UHECR air showers in the upward-facing antennas of the RNO-G detector. The identified UHECR candidate events are used to improve the knowledge about the detector and validate its performance.

\section{Description of the dataset}
\label{data-set}
\begin{figure*}
  \includegraphics[width=1\textwidth]{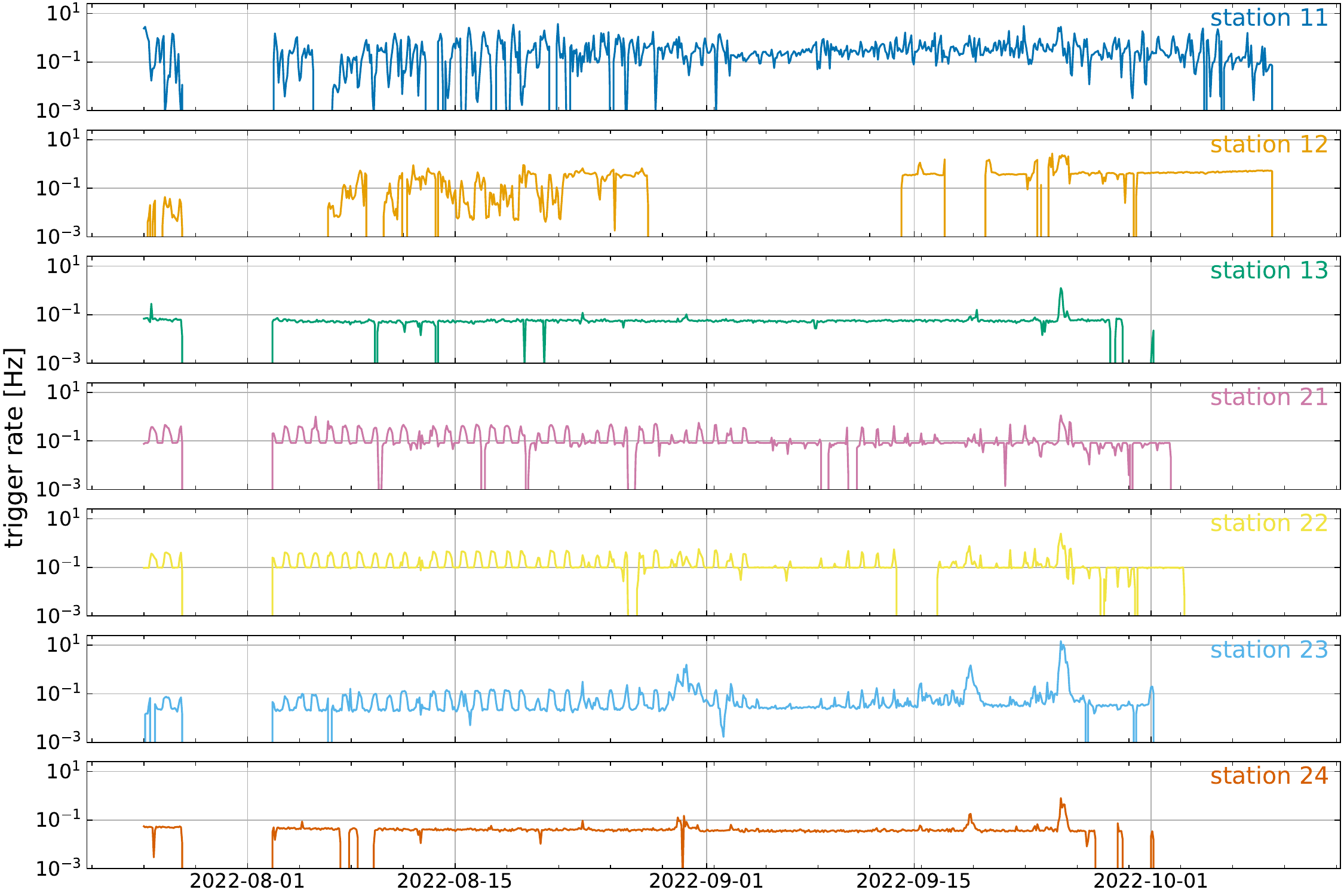}
\caption{The figure shows the trigger rate (the sum of upward and downward-facing triggers using the LPDAs) as a function of time for all science data of the seven stations in 2022.}
\label{fig:trigger_rates_2022}
\end{figure*}

\begin{figure*}
  \includegraphics[width=1\textwidth]{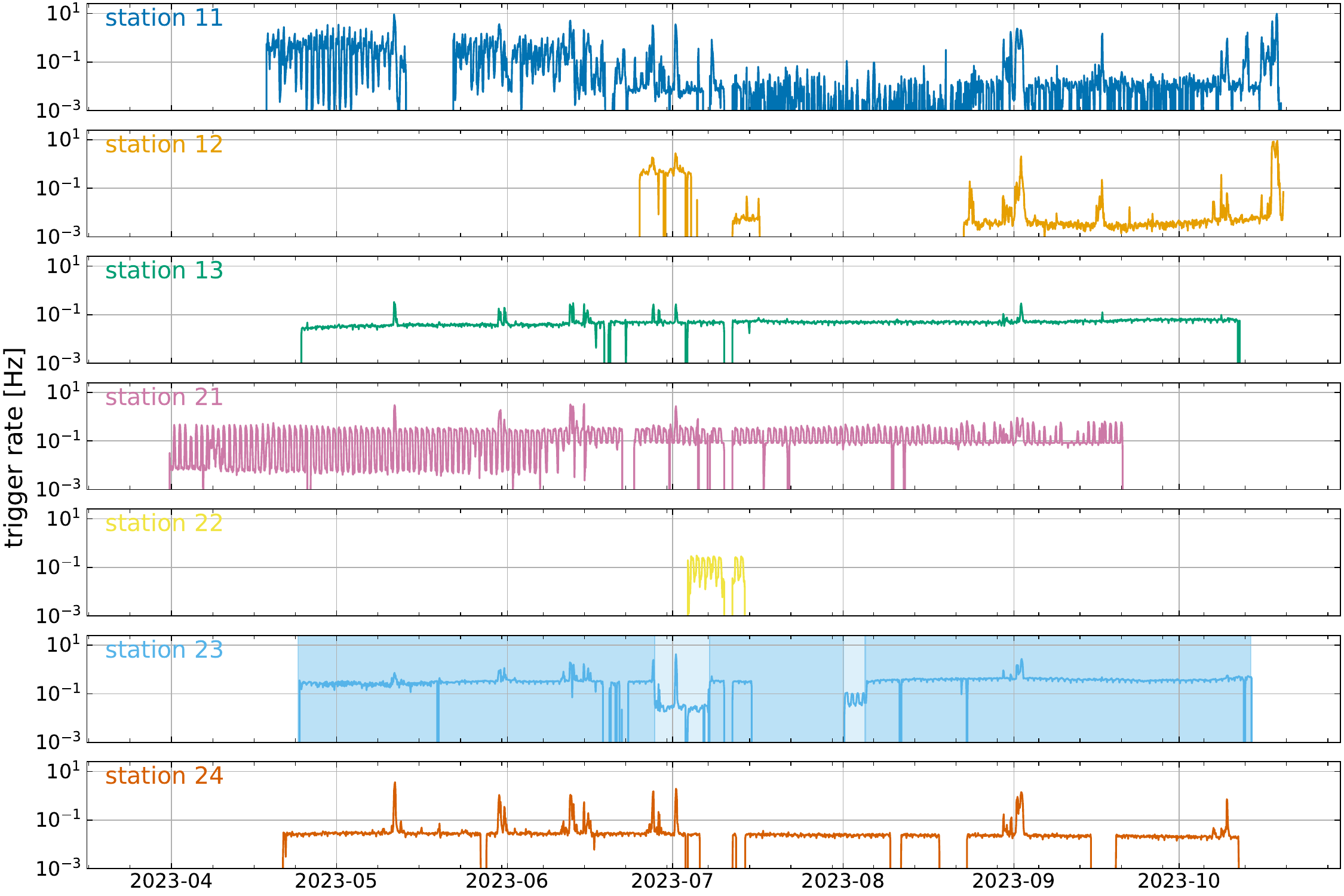}
\caption{The same as Fig.~\ref{fig:trigger_rates_2022} for the year 2023. The shaded regions for station 23 show the different configurations of that station. The first configuration is marked by the dark shaded region, while the first light shaded region marks configuration 3 and the second marks configuration 2 (see text for details).}
\label{fig:trigger_rates_2023}
\end{figure*}

For the analysis presented in this work, we consider data from 2022 and 2023, which are treated as separate datasets. Data from 2024 is excluded due to changes in the detector configuration, including modifications to the station setup at the beginning and replacement of a digitizer board during the data-taking period. During the time periods considered for this work, 7 RNO-G stations were deployed and operational. Figures \ref{fig:trigger_rates_2022} and \ref{fig:trigger_rates_2023} show the trigger rates as a function of time for the shallow antennas for all 7 stations.

Two different triggers contribute to the shallow trigger rate, both require coincident signals in multiple LPDAs. One triggers on the upward-facing LPDAs with a 2 out of 3 coincidence, while the other one triggers on the downward-facing LPDAs with a 2 out of 6 coincidence. Both triggers do not have constant trigger thresholds, but are automatically adjusted to follow the variable noise level. However, these adjustments are done per antenna and thus the resulting coincidence trigger rate can still have a non-constant behavior. This adjustment is done to avoid saturating the data-rate during periods of increased noise. Although generally the upward-facing antennas are more sensitive to air shower signals, certain event geometries result in downward-facing antennas fulfilling the trigger condition before the upward-facing ones. Since only the first trigger is recorded, both triggers are relevant for this analysis. Events from other trigger conditions of RNO-G, in particular from the phased array, are not considered for this analysis.
We use the trigger rate to quantify the performance of the stations and find that stations 13 and 24 show the most stable performance (see Fig.~\ref{fig:trigger_rates_2022} and \ref{fig:trigger_rates_2023}). Although station 23 shows small day-night variations in 2022, they occur at a more sensitive trigger threshold compared to the trigger rate variations of stations 21 and 22, making it more feasible to use data from station 23. As a result, only data from the stations 13, 23, and 24 are included in this analysis. In Sec.~\ref{cosmic_ray_selection_other_stations} we will have a closer look at the data quality of the other stations and the findings will support the trigger rate motivated decision. 
The higher and unstable trigger rates of the other stations can be explained by external factors: During the considered time period, stations 11 and 12 were equipped with a prototype wind turbine, which is thought to be responsible for the increase and large fluctuations in the trigger rates due to a poorly shielded power box \cite{rnog_instrument_paper}. The large day-night variations in the trigger rate of stations 21 and 22 can be attributed to an early version of the power system used by both stations \cite{rnog_instrument_paper}, which led to an increased noise level in the shallow antennas during the charging of the batteries from the solar panels. In both cases, we were able to design improvements, and both the wind turbines and power boxes of these stations were exchanged during the 2024 season. As a result, an improvement of the data quality for these stations is expected for the data from mid-2024 onward and will benefit future analyses.

Stations 13, 23, and 24 were installed during the 2022 season and began regular data taking after a short commissioning period. Consequently, the 2022 data-taking period started in July and lasted until October, while the 2023 period lasted from April until October. The exact dates, resulting lifetimes and number of triggered events per station can be found in Table~\ref{tab:Data_taking_periods}. We used a month of data from the 2022 data-taking period for the development of the analysis. This month of data is also included in the dataset that is used for the full analysis.

\begin{table}
\caption{The table shows the exact start/end dates, lifetime and number of triggered events of the used dataset (LPDA trigger only), separately for 2022 and 2023.}
\centering
\label{tab:Data_taking_periods}
\begin{tabular}{llcr}
\hline\noalign{\smallskip}
station & dates & lifetime [h] & \# triggers\\
\noalign{\smallskip}\hline\noalign{\smallskip}
\multirow{2}{*}{13} & 2022-07-25 & \multirow{2}{*}{1410} & \multirow{2}{*}{306,869}\\
 & 2022-10-01 &  & \\
\multirow{2}{*}{23} & 2022-07-25 & \multirow{2}{*}{1453} & \multirow{2}{*}{680,902}\\
 & 2022-10-01 &  &\\
\multirow{2}{*}{24} & 2022-07-25 & \multirow{2}{*}{1352} & \multirow{2}{*}{217,516}\\
 & 2022-10-01 &  &\\
\noalign{\smallskip}\hline\noalign{\smallskip}
\multirow{2}{*}{13} & 2023-04-24 & \multirow{2}{*}{3929} & \multirow{2}{*}{726,444}\\
 & 2023-10-12 &  &\\
\multirow{2}{*}{23} & 2023-04-24 & \multirow{2}{*}{3434} & \multirow{2}{*}{4,476,619}\\
 & 2023-10-14 &  &\\
\multirow{2}{*}{24} & 2023-04-21 & \multirow{2}{*}{3604} & \multirow{2}{*}{653,415} \\
 & 2023-10-11 &  &\\
\noalign{\smallskip}\hline
\end{tabular}
\end{table}

During the 2023 data-taking period, station 23 shows a special behavior compared to the other stations. Without changing the station settings, compared to the 2022 period, the trigger rate went up (the trigger thresholds were raised automatically), leading to a decreased sensitivity of the station. Furthermore, two periods can be identified where the trigger rate drops sharply and stays at a lower level before recovering. In what follows, we separate the 2023 period for station 23 into three different configurations and treat them separately for the analysis and simulation. The first configuration (config 1) covers everything not included in the other configurations, which is the longest period ($3119 \units{h}$ lifetime) and can be associated with the standard behavior for 2023. The second configuration (config 2) starts on the 2023-08-01 and lasts the shortest ($96 \units{h}$ lifetime). Without changing the run settings, the trigger rate went down (the trigger thresholds were lowered automatically), which leads to a more sensitive configuration, similar to the 2022 period. For the last configuration (config 3), that starts on the 2023-06-27, the downward-facing trigger was turned off and thus the trigger rate is a little lower than normal ($220 \units{h}$ lifetime). The downward-facing trigger was turned off to help investigate the sudden changes in the trigger rate, which are described above. The reason for the behavior of station 23 is unknown.

The main dataset is composed of waveforms from the upward-facing LPDAs, since they are generally more sensitive to air-shower signals than downward-facing antennas. However, the waveforms of the downward-facing antennas are used for a data reduction cut, and additionally thermal noise waveforms from the upward-facing channels are utilized to determine some of the background rejection cut values (see Sec.~\ref{Three_main_station_analysis}). The thermal noise waveforms are taken using a forced trigger that records a waveform every $10 \units{s}$. During the data taking of 2022 and 2023 the readout window for the deep and shallow antennas could not be aligned to capture both a signal in the LPDAs and the deep channels, due to analog delays in the system. 
Thus, the deep antennas do not provide any information for shallow triggered events and are not considered in this work. We do note that the problem with the misaligned readout window has been mitigated for all stations in 2024 and will enable the study of coincidence events with the shallow and deep antennas in future analyses.

\section{General analysis strategy and study of relevant systematic uncertainties}
\label{method_and_systematic}

In this section, we outline the general strategy to detect UHECR and the improvements compared to previous analyses. Furthermore, we present a study of the relevant systematic uncertainties that have an impact on the agreement between data and simulation.

\subsection{Analysis strategy}
\label{subsec:method}
\begin{figure}
  \includegraphics[width=0.49\textwidth]{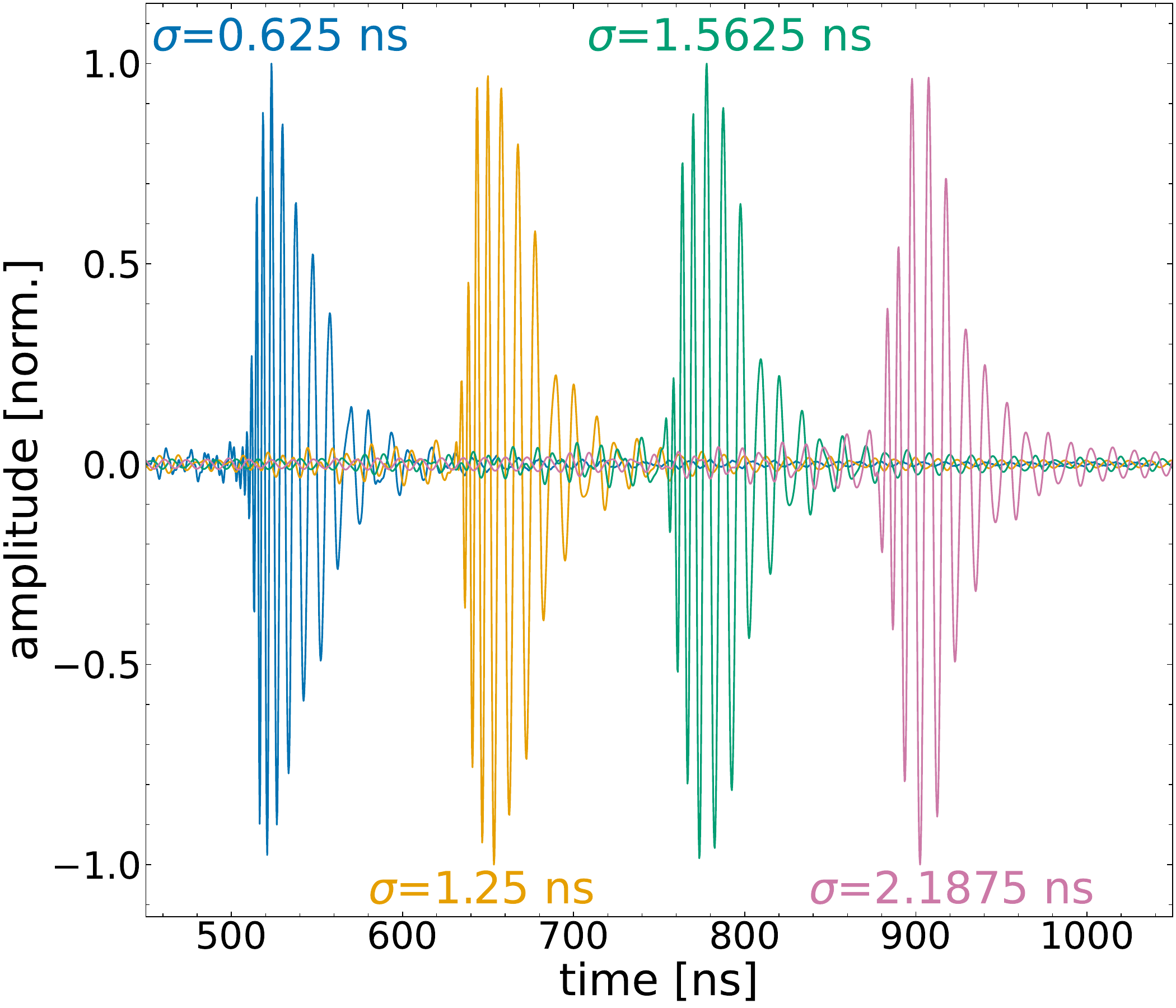}
\caption{The figure shows from left to right the waveforms of the two signal ($\sigma = 0.625\units{ns}$, $\sigma = 1.25\units{ns}$) and the two background templates ($\sigma = 1.5625\units{ns}$, $\sigma = 2.1875\units{ns}$). The Gaussian width $\sigma$ is defined in waveform samples and subsequently converted to nanoseconds, which explains the very precise values.}
\label{fig:templates}
\end{figure}

In previous analyses with the ARIANNA experiment, it was shown that a template search can be successfully applied to identify UHECR air shower signals \cite{ARIANNA_air_shower,ARIANNA_air_shower_polarization}. RNO-G uses the same LPDAs for the shallow detector as ARIANNA, therefore, we can use a similar template search to identify cosmic-ray air showers. In a template search, a set of template waveforms is correlated with the data waveforms and the resulting correlation score is used as the discriminative variable. The correlation score for which an event is considered as UHECR changes as the correlation of data and template improves with signal-to-noise ratio (SNR). Therefore, an SNR-dependent correlation score threshold is used for the UHECR event selection. However, using a large template set -- as in the initial ARIANNA analysis \cite{ARIANNA_air_shower}, which was not optimized for a small number of templates and includes $\mathcal{O}(200,000)$ template waveforms - makes this method computationally expensive, in particular for the comparatively much larger dataset of RNO-G. By exploiting the fact that the signal shape is predominantly determined by the antenna, amplification, and digitization responses and in particular by the dispersion, in a previous study we have been able to reduce the number of template waveforms to three, while still sufficiently covering the parameter space \cite{RNO-G:2023hph}. To generate the template waveforms, an artificial electric field is created. The two polarization components of this electric field are described by Gaussian functions. Then, the antenna, amplification, and digitization responses are applied to the electric field to obtain the voltage template waveform. For the antenna response, an azimuth and zenith angle needs to be assumed, for which we found that a single value for each is sufficient to cover the parameter space well enough. The width of the Gaussian function effectively changes the frequency content of the template, with a smaller width leading to more high frequencies in the final template. In the end, only two template waveforms with different frequency contents are used as the template set for the analysis. Both template waveforms, which are referred to as signal templates, are shown in Fig.~\ref{fig:templates}. Adding the third template to the template set only marginally increases the coverage by $\mathcal{O}(2\%)$ of all simulated UHECR events, at the cost of allowing more background to pass, as will be shown in Sec.~\ref{Three_main_station_analysis}. Figure~\ref{fig:templates} also shows two background-like template waveforms, which are based on Gaussian functions with a larger width, such that they match low-frequency backgrounds and not UHECR signals. 

The template search is based on a cross-correlation of signal and template. The correlation score, $\chi$, is calculated as follows:
\begin{align}
    \chi = \mathrm{max}\left(\frac{\sum_{i=0}^m(V_1)_i\cdot (V_2)_{i+\Delta n}}{\sqrt{\sum_{i=0}^m(V_1)^2_i}\cdot \sqrt{\sum_{j=\Delta n}^{m+\Delta n}(V_2)^2_j}}\right),
    \label{eq:correlation}
\end{align}
with $\mathrm{V_1}$, $\mathrm{V_2}$ being the voltage waveforms of the template and data, respectively, $\Delta n$ the number of samples by which the two waveforms are shifted relative to each other and $m$ being the number of samples of the template waveform. For the calculation of the correlation score, a scan over all possible $\Delta n$ is performed, while only the maximum of the scan is taken as the final value. Moreover, only a $200\ \mathrm{ns}$ window around the pulse of the template waveform is considered for the calculation to reduce the influence of noise. The correlation parameter, $\chi$, lies between 0 and 1, with 1 corresponding to a perfect match. Due to the normalization of the correlation parameter, the overall amplitudes of the data and template waveforms have no effect.

\subsection{Systematic uncertainties}
The effectiveness of the template search depends on how closely the template matches the UHECR waveforms. This is primarily determined by the detector's antenna and hardware (amplification and digitization) responses. Consequently, an excellent understanding of the antenna and hardware responses is crucial. Simulations indicate that even small differences in the hardware response of the search template compared to the one used for the simulation, can lead to a significant reduction of the correlation score of $\mathcal{O}(0.1)$, relative to the ideal case where both use the same response. These findings emphasize the importance of an accurate description of the detector's hardware response. In this work, a high-quality description of the hardware response is achieved by performing a full-chain measurement, using a very fast pulse generator to inject a delta-like pulse into the signal chain directly where the antenna would connect. We find that the channel-to-channel and station-to-station variance of this measurement is negligible, allowing the template set to be applied to all channels and stations.

To ensure that systematic effects beyond the hardware response are accounted for, we have investigated a variety of factors and their potential impact. A complete list of all performed checks is provided in Table~\ref{tab:systematics} together with their final impact on the correlation parameter $\chi$. Below we describe the main systematic uncertainties.

\paragraph{Index of refraction:} The uncertainty in the index of refraction $n_{\mathrm{firn}}$ of near-surface ice is the dominant systematic uncertainty in this analysis as well as potentially in all studies using data from shallow antennas. A change in $n_{\mathrm{firn}}$ slightly alters the antenna response, which directly translates into an uncertainty on the data-template correlation score if an incorrect $n_{\mathrm{firn}}$ is used for the template creation. To quantify this effect, we perform simulations in which the standard $n_{\mathrm{firn}} = 1.3$ is used for the simulated UHECR waveform, while the templates are generated with a different $n_{\mathrm{firn}}$. The antenna response at different $n_{\mathrm{firn}}$ values is obtained by rescaling the frequency ($f\rightarrow f/n_{\mathrm{firn}}$) and the gain ($\mathrm{gain} \rightarrow \mathrm{gain}\times\sqrt{n_{\mathrm{firn}}}$) of an antenna simulated in air. This scaling relation was cross-checked against an antenna simulation. Based on different ice models \cite{bob_thesis,PhysRevD.98.043010}, we estimate the range of possible $n_{\mathrm{firn}}$ at the LPDA depth to be $1.28 < n_{\mathrm{firn}} < 1.335$. Two template sets are created, using the extreme values of the estimated $n_{\mathrm{firn}}$ range for the antenna response. These modified template sets are correlated with the simulated UHECR waveforms (using the standard $n_{\mathrm{firn}}$), and SNR-dependent correlation score cut values are calculated (see Sec.~\ref{Three_main_station_analysis}) for each case. The difference with respect to the standard SNR-dependent correlation score cut is taken as the systematic uncertainty, conservatively choosing the larger difference of both cases. The resulting uncertainty is SNR dependent and as much as 0.08 in correlation score. Improvements in the understanding of the near-surface $n_{\mathrm{firn}}$ can reduce the systematic uncertainty of this analysis.

\paragraph{Antenna response:} For the antenna response, we use WIPL-D simulation of the LPDA in ice \cite{ARIANNA_air_shower}. Currently, no measurements are available that would allows us to confidently quantify variations in the response in-situ. Consequently, we estimate a systematic uncertainty due to differences in the antenna response (vector effective length) shape. To do this, the antenna response is modified by adding or removing power in different parts of the frequency spectrum, which is informed by measurements of the antenna in an anechoic chamber. For each modified response, a new set of two templates is generated. Given the large number of created sets ($\mathcal{O}(30,000)$), it is not feasible to perform simulations with each set. Instead, we correlate the templates with UHECR candidate events, found in a small subset of the data which was selected for the analysis development. The resulting differences in the correlation score compared to the standard template are taken as a guidance to estimate a conservative systematic uncertainty. The resulting uncertainty is found to be 0.015 in correlation score, which is non-negligible.

\begin{table}
\caption{The table shows a list of all considered systematic effects and their possible influence on the correlation score $\chi$.}
\centering
\label{tab:systematics}
\begin{tabular}{lS}
\hline\noalign{\smallskip}
systematic effect & impact on $\chi$ \\
\noalign{\smallskip}\hline\noalign{\smallskip}
index of refraction & < 0.08 \\
antenna response & 0.015\\
timing calibration & 0.01\\
hardware response measurement & 0.01 \\
amplifier saturation & < 0.01 \\
voltage calibration & < 0.01 \\
statistical fluctuations & < 0.01\\
station-to-station variation & < 0.01\\
galactic noise & < 0.01\\
noise temperature & < 0.01\\
antenna depth \& rotation (simulation) & < 0.01\\
\noalign{\smallskip}\hline\noalign{\smallskip}
\textbf{total} & < 0.083 \\
\noalign{\smallskip}\hline
\end{tabular}
\end{table}

\paragraph{Timing calibration:} The waveforms are digitized using a custom chip (ASIC) exploiting a switch-capacitor-array (more information can be found in Ref.~\cite{LAB4D}). To record meaningful waveforms with the digitizer, it is necessary to tune the charging times of each of the capacitors in the switch-capacitor-array. The tuning is a non-trivial and time-consuming process, and the resulting charging times can be worse than expected from the nominal performance of the chip. However, the templates are created with the assumption of a perfect digitization, which leads to a lower than expected correlation score. We investigate the effect on the correlation score by simulating waveforms that are sampled with a poorly tuned digitizer and correlate them with the perfectly sampled waveforms. For realistic charging time values, a  sub-dominant but non-negligible effect of 0.01 in correlation score was found.

\paragraph{Hardware response measurement:} A high-quality characterization of the influence of the hardware response on the signal shape is achieved by measuring the full chain with an impulsive signal. However, factors such as thermal noise, temperature or the overall stability of the measurement setup can have an influence on the precision. To assess their impact on the correlation score, we repeated the hardware response measurement at both the same (room temperature) and different temperatures (0 and $-5 ^{\circ}\mathrm{C}$). For each of the measured responses we generate a template set, which is correlated with UHECR candidate events found in a subset of data that is used for the analysis development. From the difference in the correlation score relative to the standard templates, we find an inherent uncertainty of 0.01 in correlation score that is associated to the hardware response measurements.

\paragraph{} We find that only these first four systematic effects contribute with more than or equal to 0.01 in correlation score. In the following, we will only briefly describe additional systematics that have been investigated.

\emph{Amplifier saturation:} 
We analyzed whether a potential non-linearity of the amplifiers at high amplitudes would negatively affect the signal shape and, consequentially the correlation score. Laboratory measurements showed no significant effect.

\emph{Voltage calibration:} We investigated whether a potential non-linearity in the ADC-to-voltage calibration could negatively affect the correlation score. Comparison of data-template correlation scores from waveforms with and without voltage calibration showed no significant difference.

\emph{Statistical fluctuations:} We analyzed the impact of random noise fluctuations by repeating the simulation multiple times with different noise realizations and found no significant effect.

\emph{Station-to-Station variation:} We performed the full-chain hardware response measurement for different station hardware to assess possible station-to-station variations. No significant differences between the measured responses were found.

\emph{Galactic noise:} The Galaxy is known to emit radio waves that contribute to the noise measured with the upward-facing LPDAs and thus must be included in the simulations. All simulations are created for the same day and therefore do not account for possible seasonal variations in the Galactic noise. We studied whether those seasonal variation impacted the simulations and found no significant effect.

\emph{Noise temperature:} We analyzed the impact of different thermal noise temperatures on the simulation. A change in the noise temperature effectively acts as a shift in the trigger threshold as function of SNR. Consequently, the noise temperature affects the lower SNR cutoff of the simulations and thus primarily is important for the number estimate of expected UHECR events.

\emph{Antenna position:} We studied whether a change in the antenna depth, a $180^\circ$ rotation of individual antennas around their vertical axis, or a small global rotation of the entire detector affects the simulation outcome. In all cases no significant effect was found.

\emph{Additional checks:}
We ensured that the reflection and refraction at the ice-air boundary are accurately described in the simulations. Furthermore, birefringence is expected to be negligible due to the short propagation distances $\mathcal{O}(1\units{m})$ of the signal in ice.

\section{Event selection and validation}
In this section, we describe the full analysis and simulation setup, present the results for the main dataset (stations 13, 23, and 24), report on conducted background cross-checks and reconstruct the parameters for the identified UHECR candidates. Furthermore, we assess the data quality of the stations excluded from the main dataset and provide an insight about the expected performance for the stations after the 2024 improvements.

\subsection{Simulation framework}
\label{simulation_framework}
For the simulation, we use 384 air showers that are generated using \textsc{CORSIKA} (version 7.6400) \cite{CORSIKA} and \textsc{CoREAS} (version V1.3.) \cite{Huege:2013vt}. The air showers are generated for the conditions at Summit Station and span the zenith angles $5-75^\circ$ and energies $10^{16} - 10^{18.5}\units{eV}$. Large zenith angles are excluded, since the signals have to refract into the ice to be detected, which suppresses RNO-G's sensitivity to near horizontal air showers at the surface. All showers are simulated for the same azimuth angle, but due to the nearly vertical earth magnetic field at Summit Station (inclination $80.94^\circ$ downwards) \cite{magentic_field}, we can mimic different azimuth angles by rotating the detector without introducing a large bias \cite{SJOERD-THESIS}. To span the whole parameter space, we rotate the detector in $10^\circ$ steps, and for every step, we sample each shower footprint at 240 points distributed on a star-shaped pattern. As a result, $3,317,760$ realistic events are sampled from the generated showers for the simulation. For each event that fulfills the assumed trigger condition, the correlation score is calculated and is binned as a function of SNR value to obtain the discriminative correlation score to be used on the data. 
 
\begin{figure}
  \includegraphics[width=0.5\textwidth]{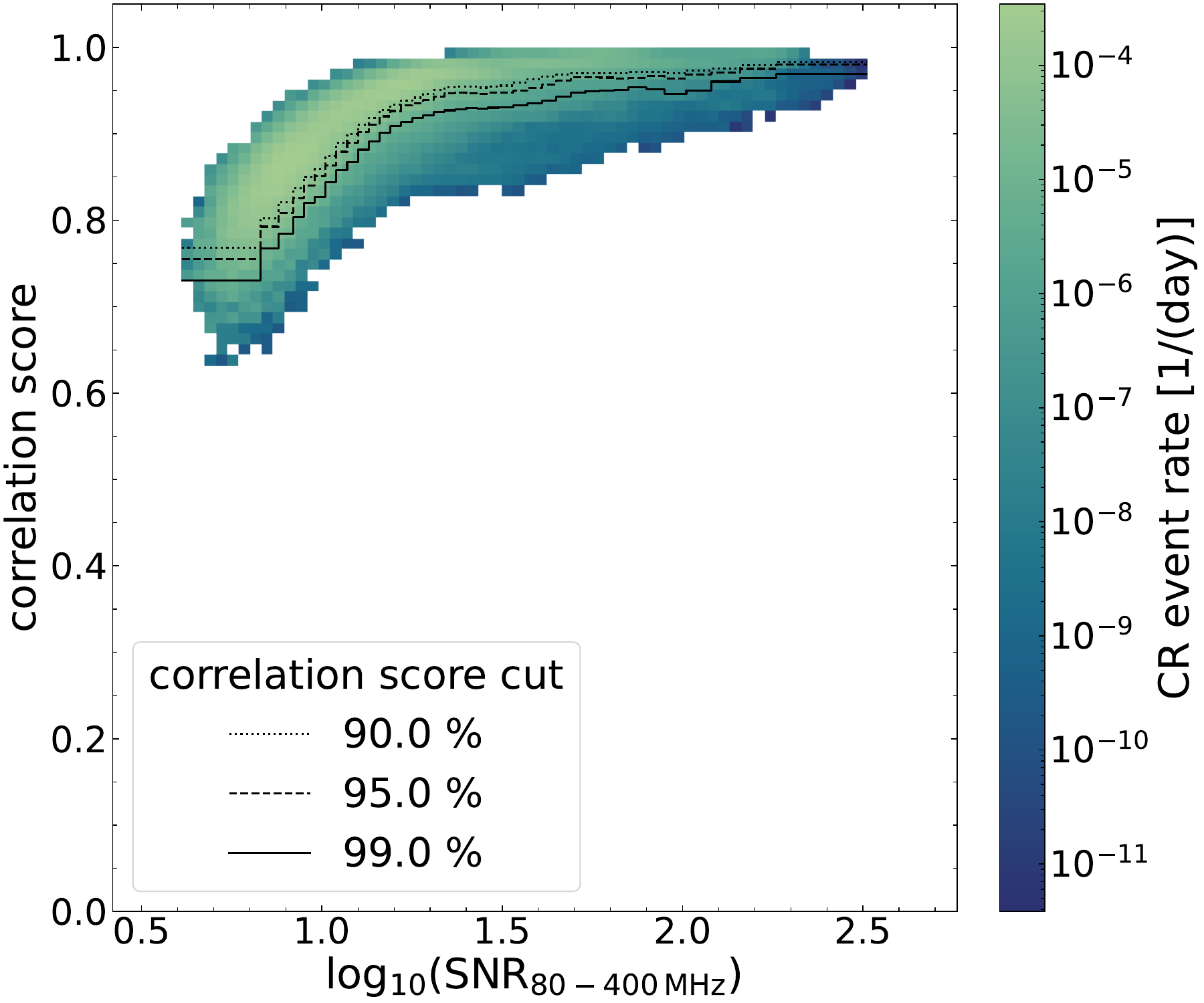}
\caption{The figure shows the correlation score as function of SNR ratio obtained from an example UHECR simulation. The colorbar indicates logarithmically the expected event rate per day. 
Also shown are three lines that illustrate the value per SNR-bin that retains 90\%, 95\%, and 99\% of the simulated UHECRs, respectively (see Sec. ~\ref{Three_main_station_analysis}). The simulation is performed with an example threshold of $25 \units{mV}$.}
\label{fig:simulation_with_analysis_cut_line}
\end{figure}

To estimate the number of expected UHECR events from simulations, we need a realistic description of the trigger. Since the real trigger is only approximated in the simulation, the threshold measured per channel on the digitizer board $\mathrm{\tau_{measured}}$ is not equal to the simulation threshold $\mathrm{\tau_{simulation}}$ and thus we need to create a mapping between both. We create the mapping by comparing trigger efficiency measurements taken in the laboratory with simulated ones. The process is described in more detail below.
As already mentioned in Sec.~\ref{data-set}, RNO-G has no fixed trigger threshold, but the threshold is dynamically adjusted to the noise level, which is influenced both by background noise and temperature. Therefore, we map per time period the mean $\mathrm{\tau_{measured}}$ to a mean $\mathrm{\tau_{simulation}}$. To estimate the mapping we used a spare digitizer board, to perform measurements in the laboratory. For different channels on this spare digitizer board, we performed trigger efficiency measurement at fixed thresholds by injecting UHECR-like waveforms into the digitizer. From these measurements, we derive a mapping between $\mathrm{\tau_{measured}}$ and the SNR value at which the trigger efficiency reaches 50\% ($\mathrm{SNR_{50\, lab}}$). The dominant uncertainty in this procedure comes from the channel-to-channel variations. Additionally, we create the same relation, mapping $\mathrm{\tau_{simulation}}$ to $\mathrm{SNR_{50\, sim}}$, by simulating trigger efficiency curves using UHECR-like waveforms. For this procedure, the dominant uncertainty comes from the difference due to the used UHECR-like waveforms. By matching $\mathrm{SNR_{50\, lab}}$ to $\mathrm{SNR_{50\, sim}}$, we translate the mean $\mathrm{\tau_{measured}}$ per channel into a simulation threshold, which allows us to run realistic simulations with an upward and downward-facing trigger. Given the large uncertainties of the resulting simulation threshold, which can be as high as $\mathcal{O}(50\%)$, we report both a best- and worst-case simulation trigger threshold. 

For station 23, the measured trigger threshold would be translated into an unrealistically small simulation threshold for the best case scenario, due to the large uncertainties from the channel-to-channel variation and extrapolation. Therefore, for that station, we use a simulation threshold of $25\units{mV}$ for all channels to calculate the SNR dependent cut on the correlation score and only provide a lower limit for the number of expected UHECR events. The threshold of $25\units{mV}$, which corresponds to $9 \sigma$, where $\sigma$ is the RMS of the simulated noise in the trigger frequency band $(80 - 180 \units{MHz})$, is lower than the best-case thresholds for the other stations, but not unrealistic. For stations 13 and 24, this problem does not occur, and all measured trigger thresholds can be mapped to realistic simulation thresholds.

Figure~\ref{fig:simulation_with_analysis_cut_line} shows an example simulation, which was performed with a constant trigger threshold of $25\units{mV}$, to illustrate the simulation procedure.

\subsection{Analysis of main dataset}
\label{Three_main_station_analysis}
\begin{figure*}
  \includegraphics[width=1\textwidth]{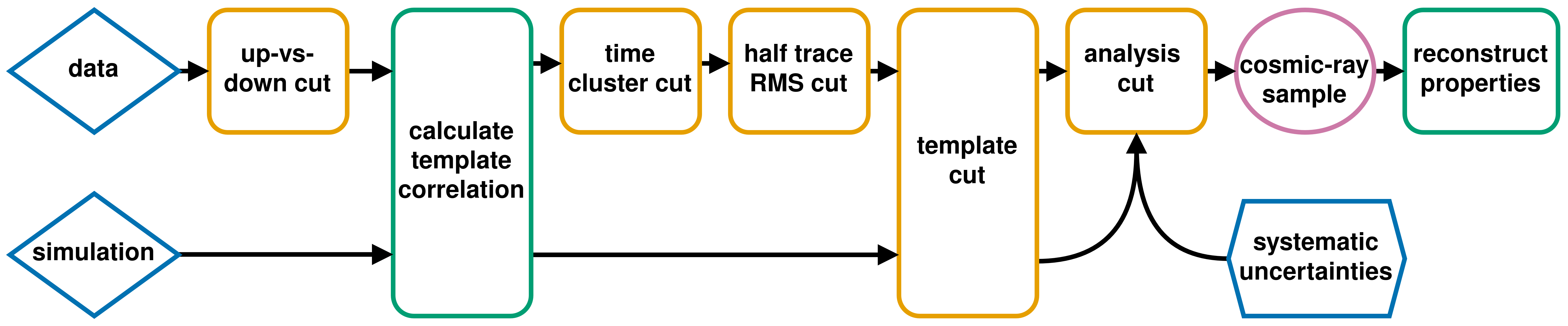}
\caption{The figure provides a schematic overview of the analysis and simulation pipeline to identify and reconstruct cosmic rays. The large blocks indicate that the same procedure is applied to both the data and simulation pipeline.}
\label{fig:analysis_simulation_pipeline}
\end{figure*}

The datasets of the three stations (13, 23, and 24) are analyzed independently while following the same analysis and simulations procedures as shown in Fig.~\ref{fig:analysis_simulation_pipeline}. In the following, we describe the analysis and simulation procedure step by step. 

For each station, the input for the analysis is a dataset as defined in Sec.~\ref{data-set}. Table~\ref{tab:event_numbers} shows the number of triggered events per station before any cuts are applied and for those remaining after the cuts that are discussed here.

\begin{table}
\caption{The table summarizes the number of events remaining after each step of the analysis for the three stations 13, 23, and 24 in the main dataset, separated by the year 2022 (top) and 2023 (bottom).}
\label{tab:event_numbers}
\begin{tabular}{lrrr}
\hline\noalign{\smallskip}
  & 13 & 23 & 24 \\
\noalign{\smallskip}\hline\noalign{\smallskip}
before all cuts & 306,869 & 680,902 & 217,516 \\
after up-vs-down cut & 115,134 & 547,083 & 63,377 \\
after time cluster cut & 100,983 & 103,329 & 38,606 \\
after half trace RMS cut & 97,374 & 99,193 & 37,375 \\
after template cut & 82,366 & 86,342 & 32,489 \\
\noalign{\smallskip}\hline\noalign{\smallskip}
before all cuts & 726,444 & 4,476,619 & 653,415 \\
after up-vs-down cut & 201,867 & 1,263,333 & 375,271 \\
after time cluster cut & 163,287 & 769,029 & 65,081 \\
after half trace RMS cut & 151,590 & 752,959 & 61,263 \\
after template cut & 129,768 & 652,425 & 52,273 \\
\noalign{\smallskip}\hline
\end{tabular}
\end{table}

In the first analysis step, the \emph{up-vs-down cut} is applied, removing all events from the dataset in which the amplitude in the downward-facing antennas is higher than in the upward-facing. This cut ensures that the triggered signal comes from above the ice. Such a signal arrives at the downward-facing antenna in the insensitive back-lobe, resulting in a smaller signal amplitude compared to the upward-facing antennas, where the signal arrives in the sensitive front-lobe. Primarily, thermal noise events are rejected by this cut. Its main purpose is to reduce the number of events in the dataset, thereby speeding up the following steps. The number of events surviving the up-vs-down cut are shown in Table~\ref{tab:event_numbers}. 

In the next step, the data-template correlation is calculated (\emph{template correlation}). For each event, the waveforms from all upward-facing channels are correlated with all four templates (two signal and two background templates) using Eq.~\ref{eq:correlation}, resulting in 12 correlation scores per event. Only the maximum of these 12 correlation scores is considered in the next steps of the analysis. Two additional important event quantities are then calculated from the channel waveform with the highest correlation score. The first is the half trace RMS (htRMS) value, defined as the root-mean-square (RMS) of the first half of the waveform, which is in ADC counts. The second value is the SNR, defined as the ratio of the maximal amplitude and the htRMS value.  

Subsequently, a series of background rejection cuts are applied to the data. First, the \emph{time cluster cut} is applied which is designed to remove backgrounds that cluster in time and have a high correlation score. This cut is effective because UHECR events are randomly distributed in time rather than clustered. Possible backgrounds are, however, temporally clustered such as signals from airplanes \cite{airplanes} and background signal occurring during high wind periods \cite{triboelectric}. While these backgrounds are likely to be identifiable by their signal shape (e.g.~\cite{ARIANNA_air_shower}), without a model for these backgrounds, we choose the conservative approach to use a time cluster cut. To minimize the number of random coincidences from low correlation score events that are not signal-like, only events that have a correlation score above a certain value (minimal correlation score) are considered for this cut. An event is retained if no other event with a correlation score larger than the minimal correlation score occurs within the time period $\Delta \mathrm{t}$. If two or more events are within $\Delta \mathrm{t}$ of each other, the whole time period is rejected from the dataset, including those events that did not contribute to the cut due to their small correlation score. 

\begin{figure}
  \includegraphics[width=0.5\textwidth]{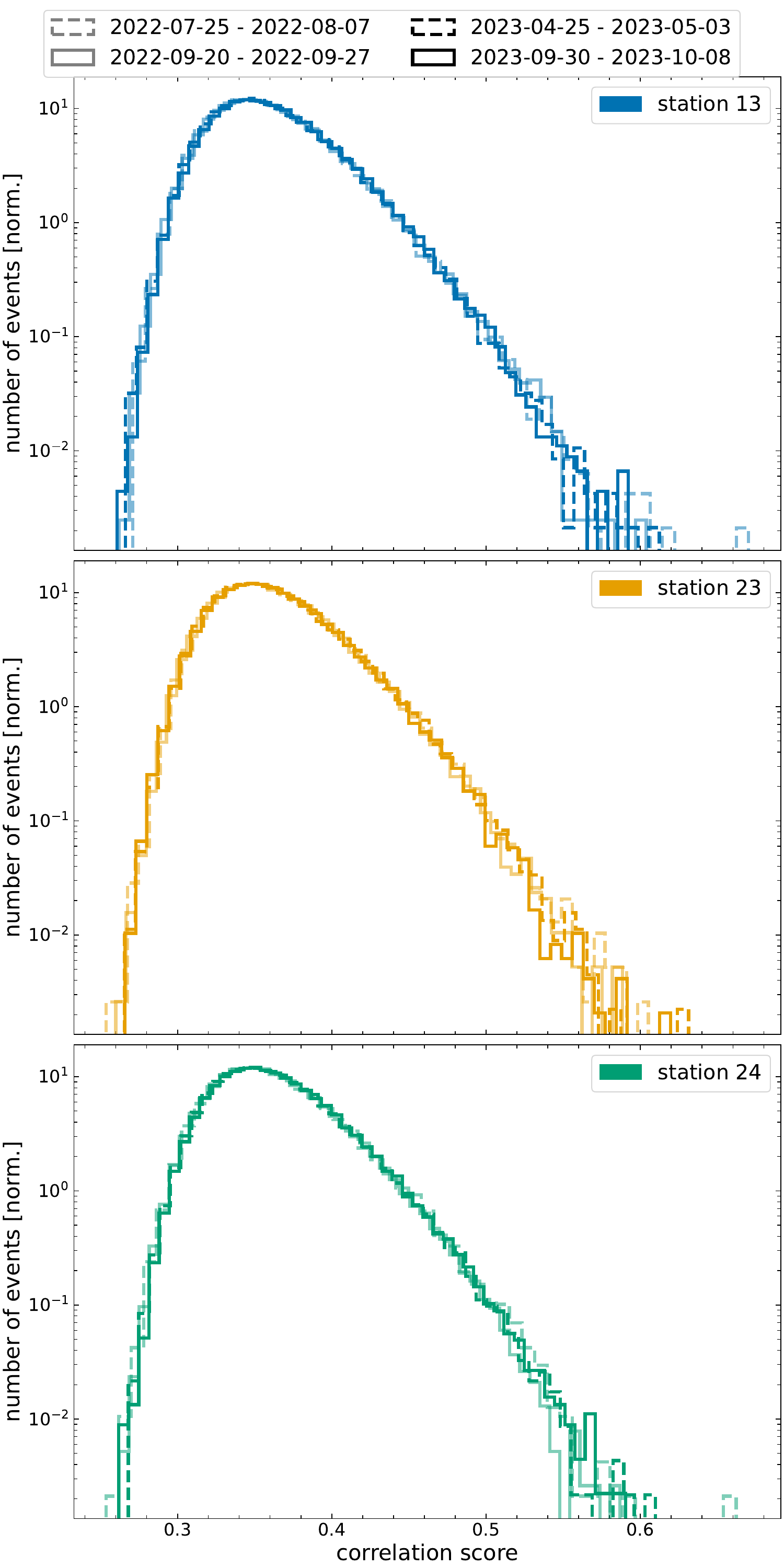}
\caption{The distribution of the correlation scores of forced trigger data for three stations during different time periods in 2022 and 2023.}
\label{fig:time_cluster_cut_min_corr_value}
\end{figure}

We determined the value for the time period $\Delta\mathrm{t}$ by optimizing it on a small subset of the data that is used for analysis development. For all three stations we use a value of $\Delta\mathrm{t}=60\units{s}$. We derive the minimal correlation score from the correlation score distribution of thermal noise events which are taken following a forced trigger. Figure~\ref{fig:time_cluster_cut_min_corr_value} shows these distributions for all stations and both data-taking periods (2022 and 2023). The figure shows, for both years, the distribution at the beginning and the end of the data-taking period. All four distributions are stable over time, which makes it possible to choose a constant cut value.
For each station, the value is chosen such that only a few thermal noise events would contribute to the cut and thus reduce the number of possible random coincidences.  To further support the decision, a simulation was performed to estimate the expected influence of random coincidence due to thermal noise as a function of the minimal correlation cut value. Each dataset contains a realistic number of thermal noise and UHECR events, both with a random time-stamp and a correlation score above the minimal correlation score. From this simulation, we calculate the expected lifetime loss and UHECR dead time due to random coincidences. The UHECR dead time arises because each thermal noise event with a correlation score higher than the minimal correlation score effectively blocks a time window of $2\Delta{t}$, during which no UHECR event can occur without being rejected by the time cluster cut. This effective UHECR dead time needs to be accounted for when estimating the expected number of UHECR events. 
For all three stations and both years, we set the minimal correlation score to be 0.55.
The number of events surviving the time cluster cut can be found in Table~\ref{tab:event_numbers}, which corresponds in 2022 (2023) to a lifetime loss of $0.3\%$ ($1.1 \%$) for station 13, $7.6 \%$ (config 1: $4.8 \%$, config 2: $0.02 \%$, config 3: $ 4.5\%$) for station 23 and $2.2 \%$ ($4.5 \%$) for station 24.

\begin{figure}
  \includegraphics[width=0.5\textwidth]{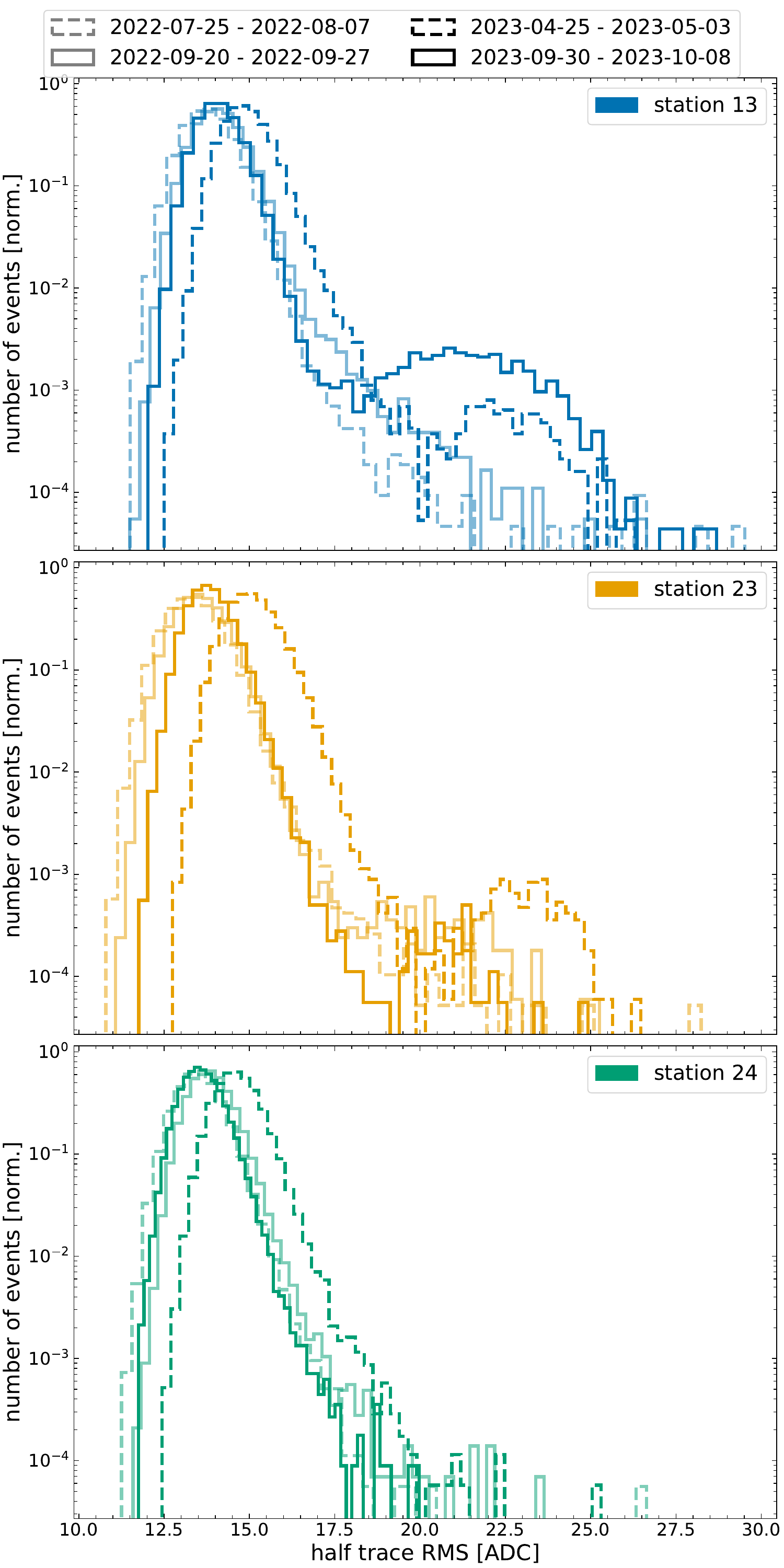}
\caption{The figure shows the half trace RMS (htRMS) distribution of forced trigger data for three stations and different time periods in 2022 and 2023.}
\label{fig:half_trace_RMS_values}
\end{figure}

As the second background rejection cut we apply the \emph{htRMS cut}. The signal is, by construction, in the second half of the recorded waveform given the trigger settings. For clean UHECR pulses, we only expect thermal noise on the first half of waveforms. Events with a signal other than thermal noise on the first half of the waveform - quantified by the htRMS value - are rejected as background. Possible background events rejected by the htRMS cut are solar flare events \cite{solar_flare}. However, this cut is fundamentally agnostic to the background's origin. 
The htRMS threshold, above which a signal is considered background, is estimated from htRMS thermal noise distributions. Figure~\ref{fig:half_trace_RMS_values} shows, for each station and year, two thermal noise htRMS distributions: one at the beginning and one at the end of the data-taking period. 

For each station, we can see in Fig.~\ref{fig:half_trace_RMS_values} that the htRMS distribution changes during the data-taking period. The changes in 2022 are relatively small compared to the ones in 2023, which likely can be attributed to the longer data-taking period in 2023. Furthermore, for some stations, some distributions have a smaller second peak at higher htRMS values, which can be attributed to narrow band events (e.g., hand-held radio communication or data transmission from a weather balloon) that make it into the forced trigger dataset. The instability of the htRMS distribution makes it necessary to define a value for the cut that is time-dependent. We calculate the time-dependent threshold by splitting the data into bins of 5 days. For each time bin, we fit a log-normal distribution to the main peak and estimate the cut value by taking the 0.99999 quantile from the fit. With this approach, it is possible to set a cut value that is not artificially increased by the waveforms with excess noise, such as narrow band events.
The number of events that survive the htRMS cut can be found in Table~\ref{tab:event_numbers}.

The last background rejection cut is the \emph{template cut}. A significant number of impulsive backgrounds with a high correlation score have a much lower frequency content compared to what is expected from UHECR events. As a result, we introduced additional back\-ground-tagging templates with a low frequency content ($\sigma=1.5625\units{ns}$, $\sigma=2.1875\units{ns}$), shown in Fig.~\ref{fig:templates}. Events that have the highest correlation score with either of the two background templates are rejected from the dataset. Simulations show that for all stations, less than $2.2\%$ of the expected UHECR events are rejected, which makes this cut very efficient. The number of events surviving the last background rejection cut are shown in Table~\ref{tab:event_numbers}.
After applying all background rejection cuts, an analysis cut is applied to selected the UHECR sample. Before introducing the analysis cut, we first describe the simulation procedure used to create it.

After generating the simulation (see Sec.~\ref{simulation_framework}), the correlation score, htRMS and SNR are calculated for all triggered events. Subsequently, the background rejection cuts are applied. Since this is pure UHECR simulation, only the template cut has an effect, as the time-cluster and htRMS cut do not remove any simulated events. It should be noted that also the up-vs-down cut is not applied, as it does not affect the pure UHECR simulation.

Finally, the simulation is used to calculate the SNR-dependent cut on the correlation score to select the sample of UHECR candidate events. Figure~\ref{fig:simulation_with_analysis_cut_line} shows three lines that are calculated from the shown simulation. The interpretation of each line is that 90\%, 95\%, and 99\% of all simulated UHECR events lie above the line in correlation score per SNR-bin.
To determine these lines, the simulation is divided into multiple SNR bins, each containing roughly $10,000$ events to ensure sufficient statistics. For each SNR bin, the correlation cut value is chosen such that the desired fraction of simulated UHECR survives the cut. To ensure that all possible UHECR events are selected, the SNR-dependent correlation score is calculated using the best-case instead of the worst-case simulation (for station 23, a simulation with a $25 \units{mV}$ threshold has been be used). As discussed in Sec.~\ref{method_and_systematic}, four systematic uncertainties significantly influence the correlation score (larger than 0.01 in correlation score). These systematic uncertainties are added quadratically for each bin. Except for the uncertainty from the hardware response measurement, all systematic uncertainties consistently lower the correlation score per bin. The uncertainty from the hardware response can also increase the correlation score per bin, leading to an asymmetric systematic uncertainty region.

\begin{figure*}
  \includegraphics[width=1\textwidth]{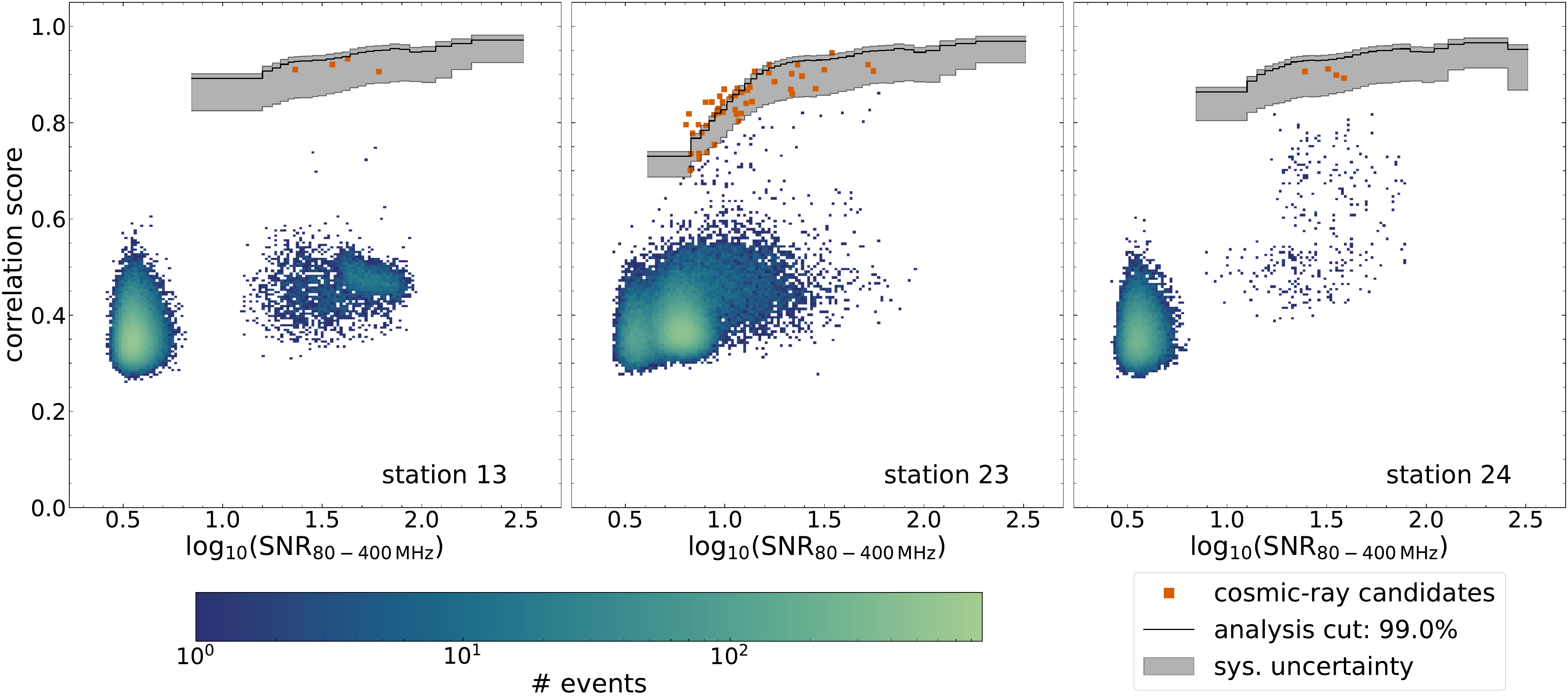}
\caption{The figure shows the correlation score as function of SNR for the complete 2022 dataset per station after applying all cuts. The selected UHECR candidate events are highlighted in orange.}
\label{fig:2022_data}
\end{figure*}

\begin{figure*}
  \includegraphics[width=1\textwidth]{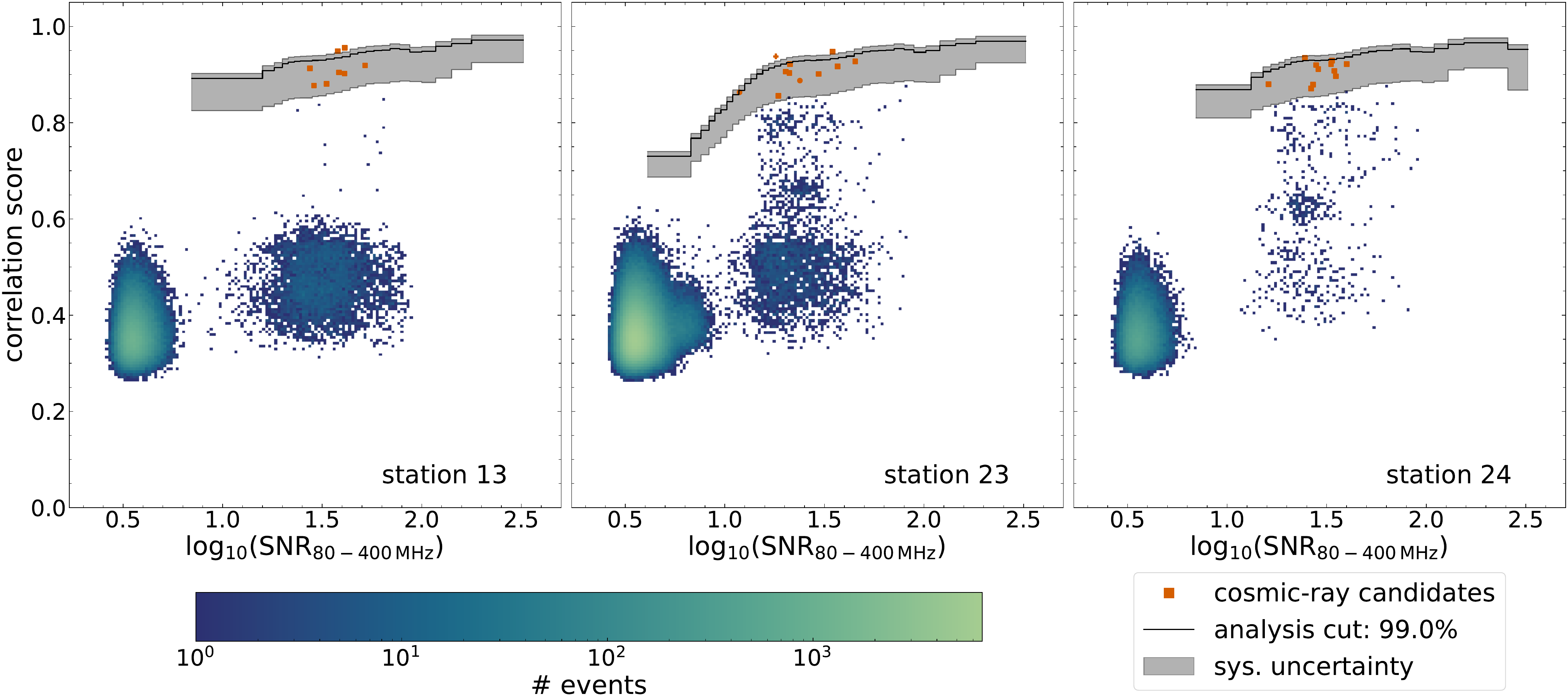}
\caption{Same as Fig.~\ref{fig:2022_data} for 2023. For station 23, only the most conservative SNR-dependent correlation score cut is shown and different markers indicate the candidate events from different configurations.}
\label{fig:2023_data}
\end{figure*}

The determined SNR-dependent correlation score cut is applied as the final \emph{analysis cut} to the data. 
The dataset for all three stations, after applying all background rejection cuts, is shown for 2022 in Fig.~\ref{fig:2022_data} and for 2023 in Fig.~\ref{fig:2023_data}. Also shown in Fig.~\ref{fig:2022_data} and \ref{fig:2023_data}, is the calculated SNR-dependent cut on the correlation score, chosen to retain 99\% of the simulated UHECR events (including the systematic uncertainties) and the selected sample of UHECR candidate events.

From the simulation, we obtain a number for the expected rate of UHECR events per day for the best- and worst-case scenario, based of the uncertainties stemming from the realistic simulation threshold. Multiplying this rate by the detector lifetime yields a range for the expected number of UHECR events from the dataset. We also account for the lifetime loss due to the time cluster cut, the UHECR dead time due to random coincidences from the time cluster cut, and the rate reduction due to the template and analysis cut. The expected and found number of UHECR candidate events per station and year are summarized in Table~\ref{tab:number_of_cr}. Given the conservative best- and worst-case scenario estimate, we expect the number of all found UHECR candidate events to lie within the expected range for all stations and years. For station 23, only a lower limit is available, as no best-case scenario could be calculated (see Sec.~\ref{simulation_framework}). In agreement with expectations, the number of identified UHECR candidate events lies within the predicted range for all stations and years.

\begin{table}
\caption{The table shows the found and expected number of UHECR candidate events for all stations and datasets.}
\centering
\label{tab:number_of_cr}
\begin{tabular}{lllll}
\hline\noalign{\smallskip}
& station &  &expected & found \\
\noalign{\smallskip}\hline\noalign{\smallskip}
\multirow{3}{*}{2022} & 13 & & 1 - 5 & 4  \\
& 23 & & $>$ 3 & 47 \\
& 24 & & 2 - 8 & 4 \\
\noalign{\smallskip}\hline\noalign{\smallskip}
\multirow{5}{*}{2023} & 13 & & 4 - 13 & 8  \\
& \multirow{3}{*}{23} & config 1 & $> 4$ & 8  \\
&  & config 2& $>$ 0 & 2  \\
&  & config 3 & 0 - 2 & 1  \\
& 24 & & 4 - 19 &  11 \\
\noalign{\smallskip}\hline
\end{tabular}
\end{table}

Figures~\ref{fig:2022_data} and \ref{fig:2023_data} show that the majority of the UHECR candidate events would not satisfy the correlation cut without accounting for systematic uncertainties, underscoring the importance of their proper treatment. The largest number of UHECR candidate events is found for Station 23 in the year 2022, which is consistent with the station having a lower trigger (and thus cosmic-ray energy) threshold for this year compared to the other stations. Due to the lower energy threshold and the cosmic-ray spectrum, we expected for this station to find the majority of the UHECR candidate events at lower SNR (lower energy), which is in agreement with the result shown in Fig.~\ref{fig:2022_data}. As discussed in Sec.~\ref{data-set}, the trigger threshold for station 23 was increased for most of 2023 relative to the previous year. Consequently, we expect fewer UHECR events and those detected should have a higher SNR value, which is reflected in the data.

From the left panels of Figures~\ref{fig:2022_data} and \ref{fig:2023_data}, it can be shown that two populations of background events -- one at lower and the other at higher SNR -- survive the background rejection cuts for station 13. Because of the employed noise-riding trigger threshold, only the population at higher SNR ($\log_{10}(\mathrm{SNR}) \gtrsim 1$) is expected, with the lowest SNR value of this population proportional to the station's trigger threshold. The unexpected distribution at low SNR ($\log_{10}(\mathrm{SNR}) \lesssim 1$) is consistent with the SNR and correlation score of thermal noise and its existence can be attributed to the trigger hardware. For the version of the digitizer boards used to collect the presented data, the incoming signal is split between the trigger and signal path for the shallow channels, with the majority of the power directed to the signal path. This leads to the trigger path being dominated by board noise, causing the system to trigger on this noise, which is not visible in the recorded waveforms (i.e.\ signal path). As a result, the board noise produces false positive triggers that make up the population of thermal noise events. These events pass all background rejection cuts, as no cut is specifically designed to remove thermal noise events. Also, the efficiency for cosmic rays at low SNR values is poor. 
The same behavior as for station 13 can be seen for station 24 (right panel, Figures~\ref{fig:2022_data} and ~\ref{fig:2023_data}), while it is less clear in station 23 (center panel, Figures~\ref{fig:2022_data} and \ref{fig:2023_data}), especially in 2023 due to its multiple configurations (see Sec.~\ref{data-set}). Furthermore, comparing stations 13 and 23 for the year 2022, which have different thresholds, the population at higher SNR values shifts to smaller SNR values for the station with the lower threshold (station 23), which shows the dependence on the trigger threshold for this population. For a small enough trigger threshold, the population at high SNR values and the thermal noise population overlap, as visible in the station 23 data.
The above-described issue with the digitizer board is known, and the next digitizer version (v3) features improved signal splitting to mitigate it. For all stations except 24, the digitizer board was upgraded in 2024. Studies of the trigger efficiency of the new digitizer board in a laboratory set-up have shown that we expect an increased rate of $5^{+5}_{-2}$ UHECR events per day and per seven stations \cite{LILLY-THESIS}. As a result, with the new design, in less than $3000 \units{h}$ of lifetime of a single station, the number of UHECR events is comparable to those found in this work, which uses a total of $15,182 \units{h}$ of single station lifetime. Future analyses will benefit from the improved trigger performance and the increased UHECR statistic. 

\begin{figure}
\begin{subfigure}{0.8\textwidth}
    \includegraphics[width=0.5\textwidth]{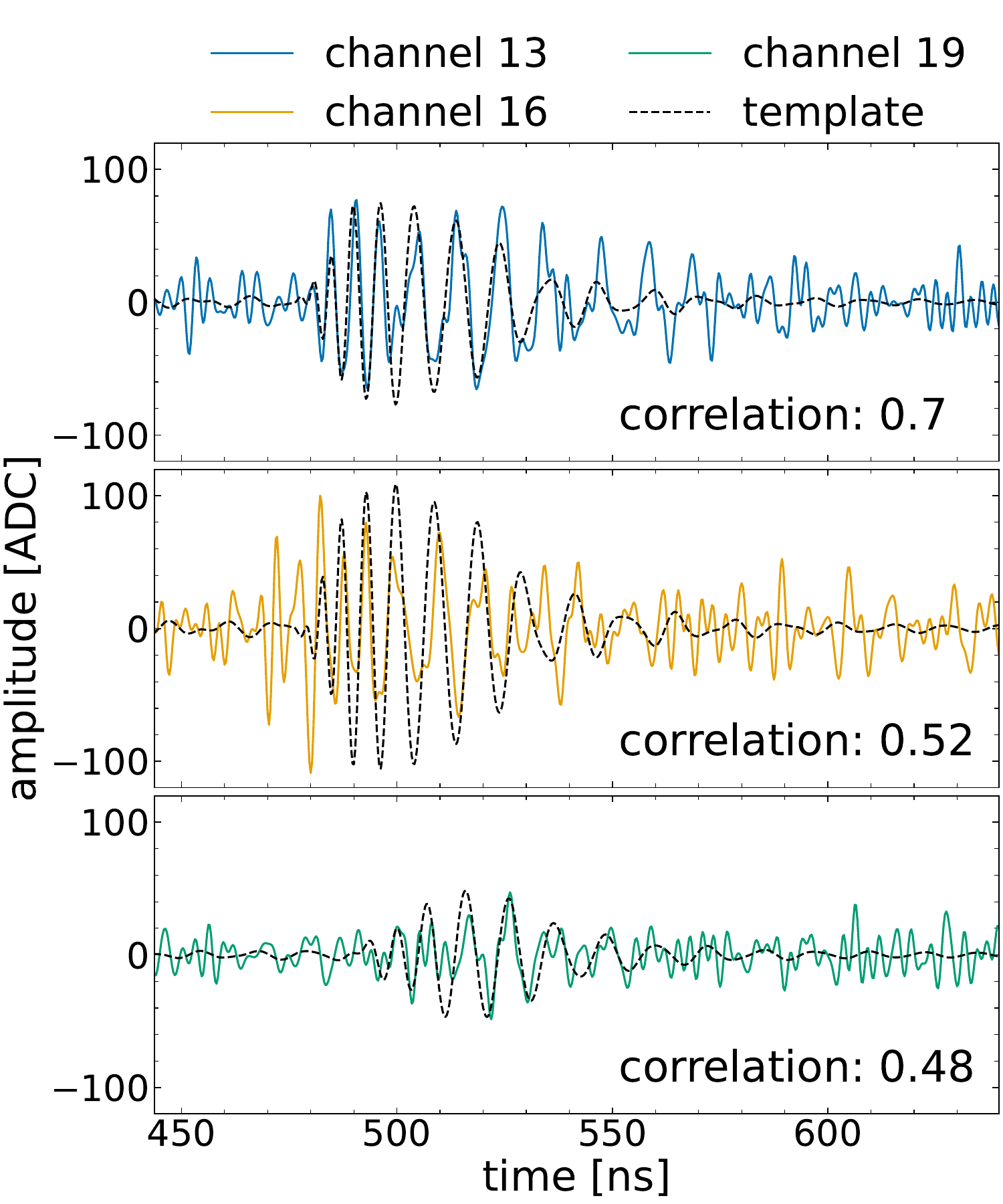}
\end{subfigure}

\begin{subfigure}{0.8\textwidth}
    \includegraphics[width=0.5\textwidth]{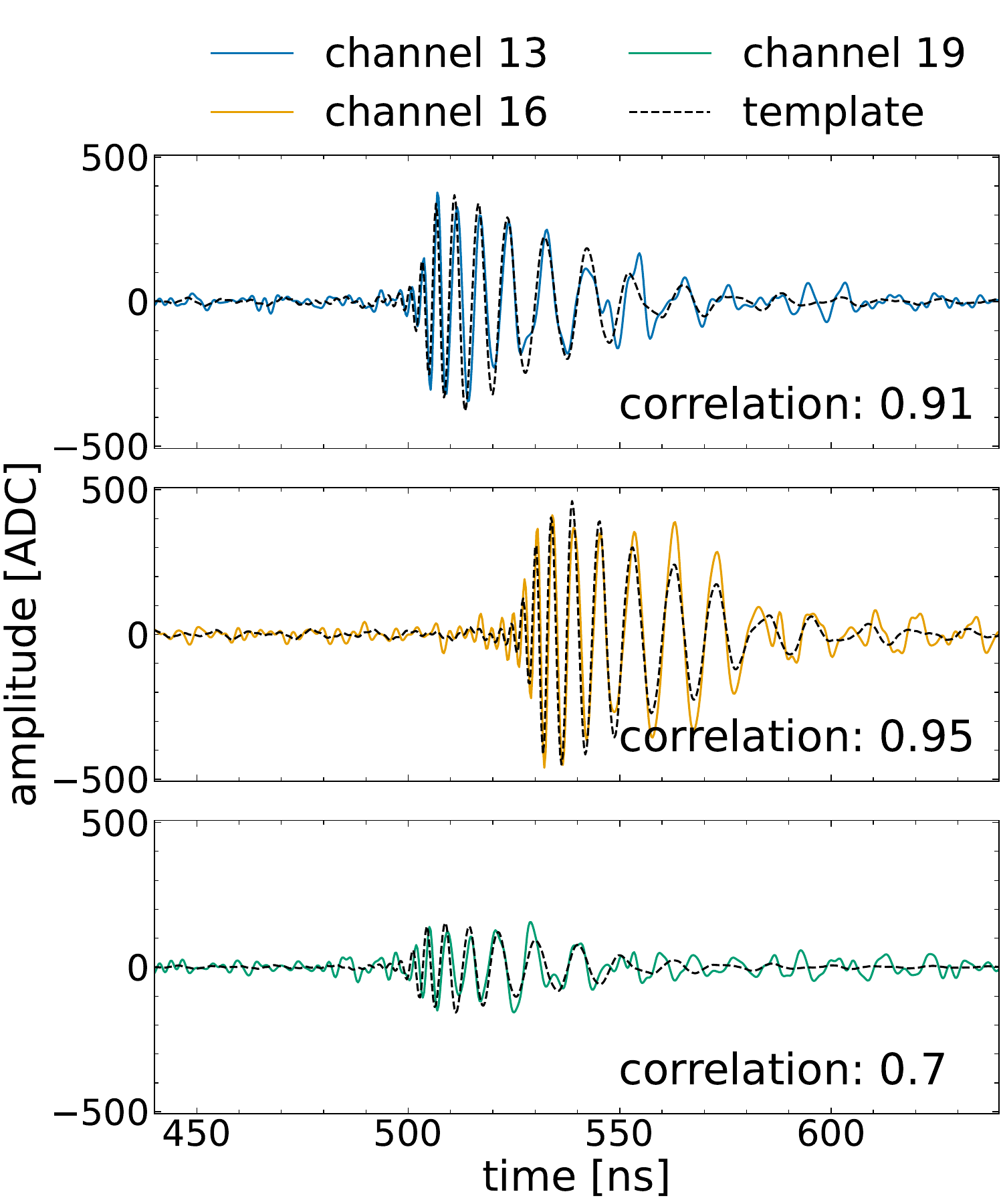}
\end{subfigure}
\caption{The figure shows the waveforms of the upward-facing LPDAs for the UHECR candidate event with the lowest (upper figure) and highest (lower figure) correlation score in the 2022 data.}
\label{fig:cr_candidate_2022}
\end{figure}

\begin{figure}
\begin{subfigure}{0.8\textwidth}
    \includegraphics[width=0.5\textwidth]{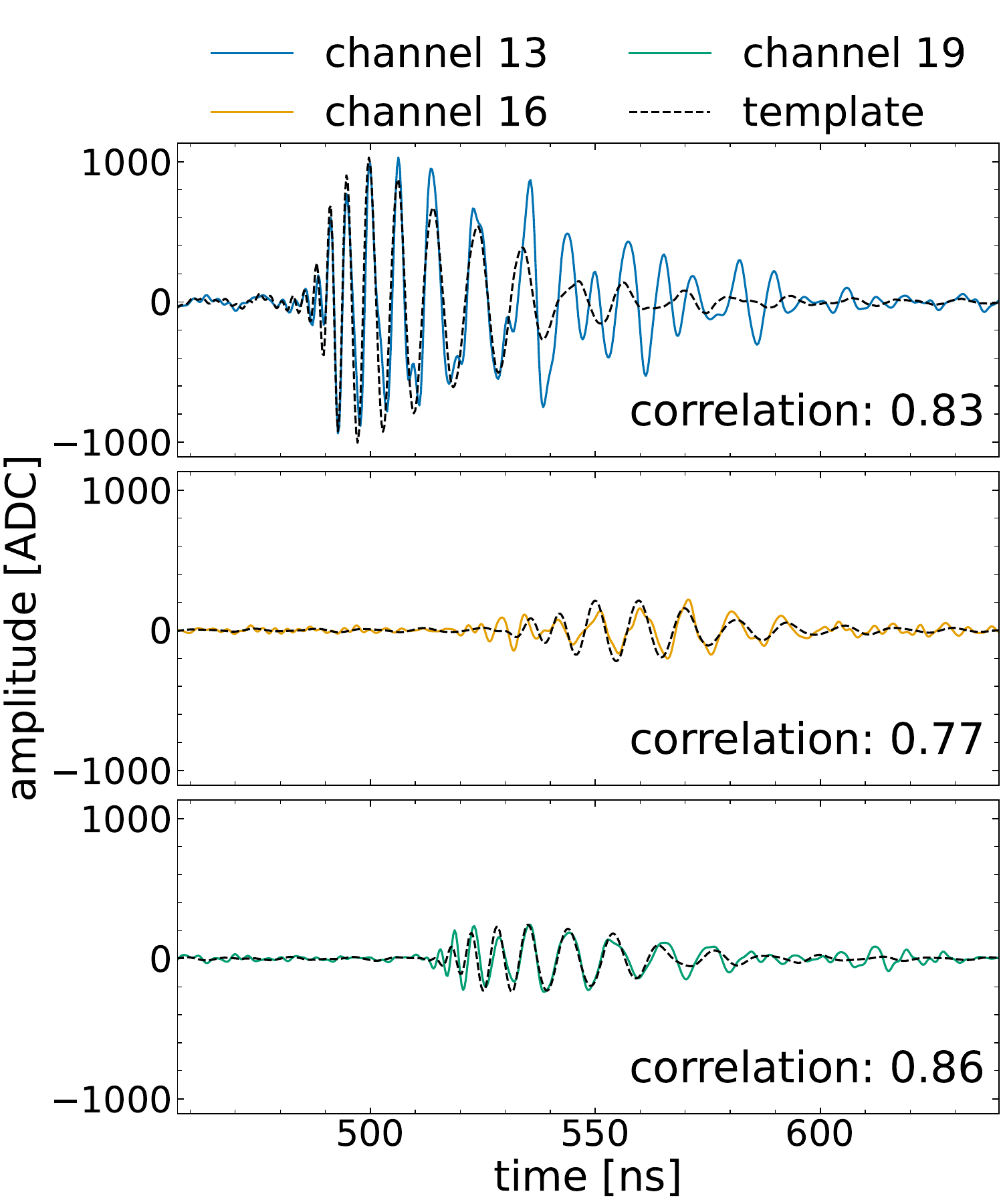}
\end{subfigure}

\begin{subfigure}{0.8\textwidth}
    \includegraphics[width=0.5\textwidth]{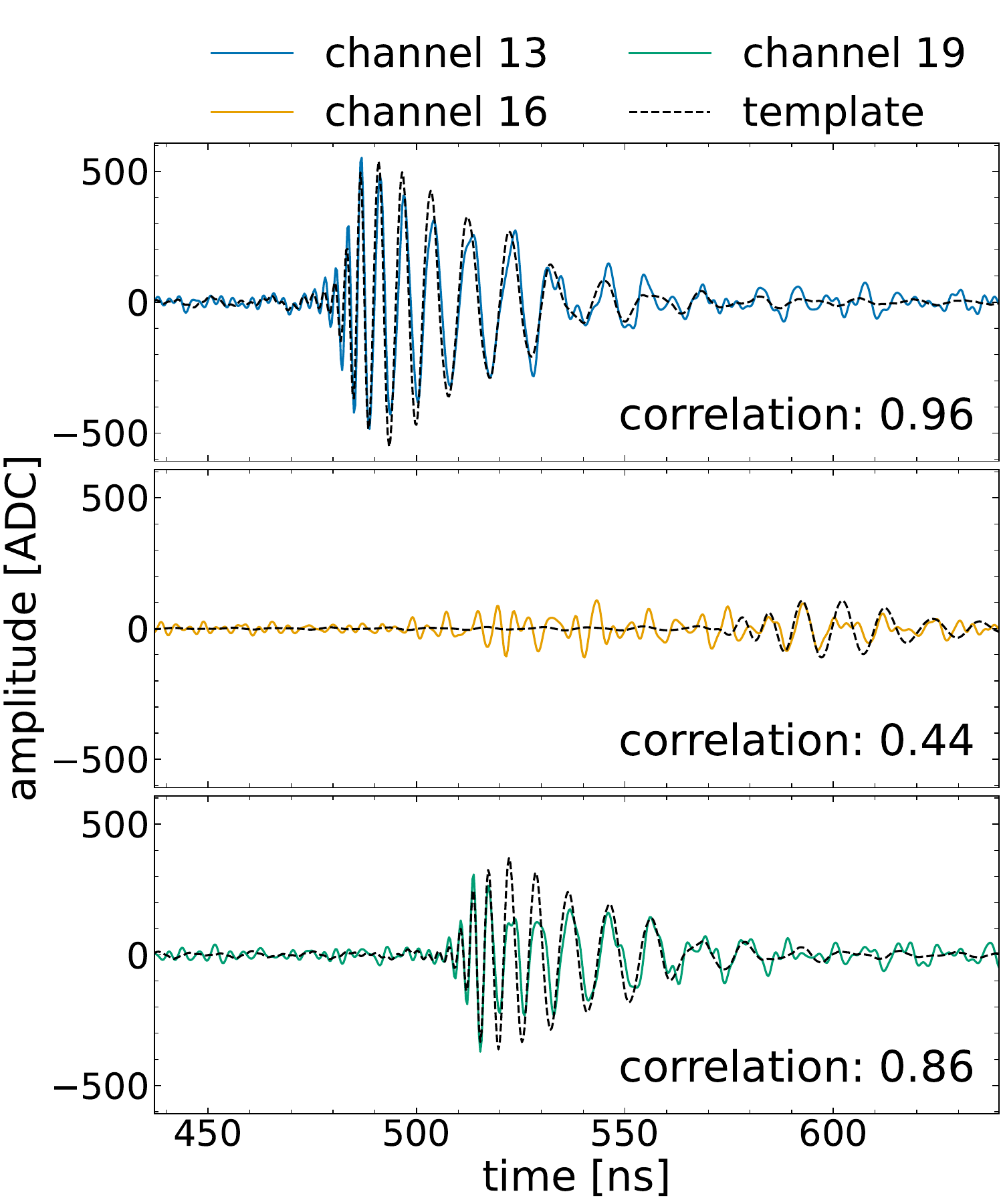}
\end{subfigure}
\caption{The figure shows the waveforms of the upward-facing LPDAs for the UHECR candidate event with the lowest (upper figure) and highest (lower figure) correlation score in the 2023 data.}
\label{fig:cr_candidate_2023}
\end{figure}

To illustrate the events selected, Fig.~\ref{fig:cr_candidate_2022} and \ref{fig:cr_candidate_2023} display for each year the waveforms of the UHECR candidates with the highest and lowest correlation score. 
For each candidate event, the waveforms of the three upward-facing channels are overlaid with the best fitting template to illustrate the data-to-template match.
The signal properties - the zenith angle, azimuth angle, polarization, and energy - for all UHECR candidates are reconstructed as will be discussed in Sec.~\ref{reconstruction_CR}.


\subsection{Cross-check for background contamination}
\label{sec:background}
Given the discussion of systematic uncertainties and the distributions shown in Fig.~\ref{fig:2022_data} and \ref{fig:2023_data}, it is unclear whether a perfect separation of signal and background is possible. We checked that the cosmic-ray candidates are unlikely to stem from the known background of airplanes \cite{airplanes}. Furthermore, we checked that the candidates show a distribution that is flat with respect to the time of day and do not correlate with periods of high winds, which is believed to cause most of the signal-like backgrounds \cite{triboelectric}.

\begin{figure}
    \centering
    \includegraphics[width=0.95\linewidth]{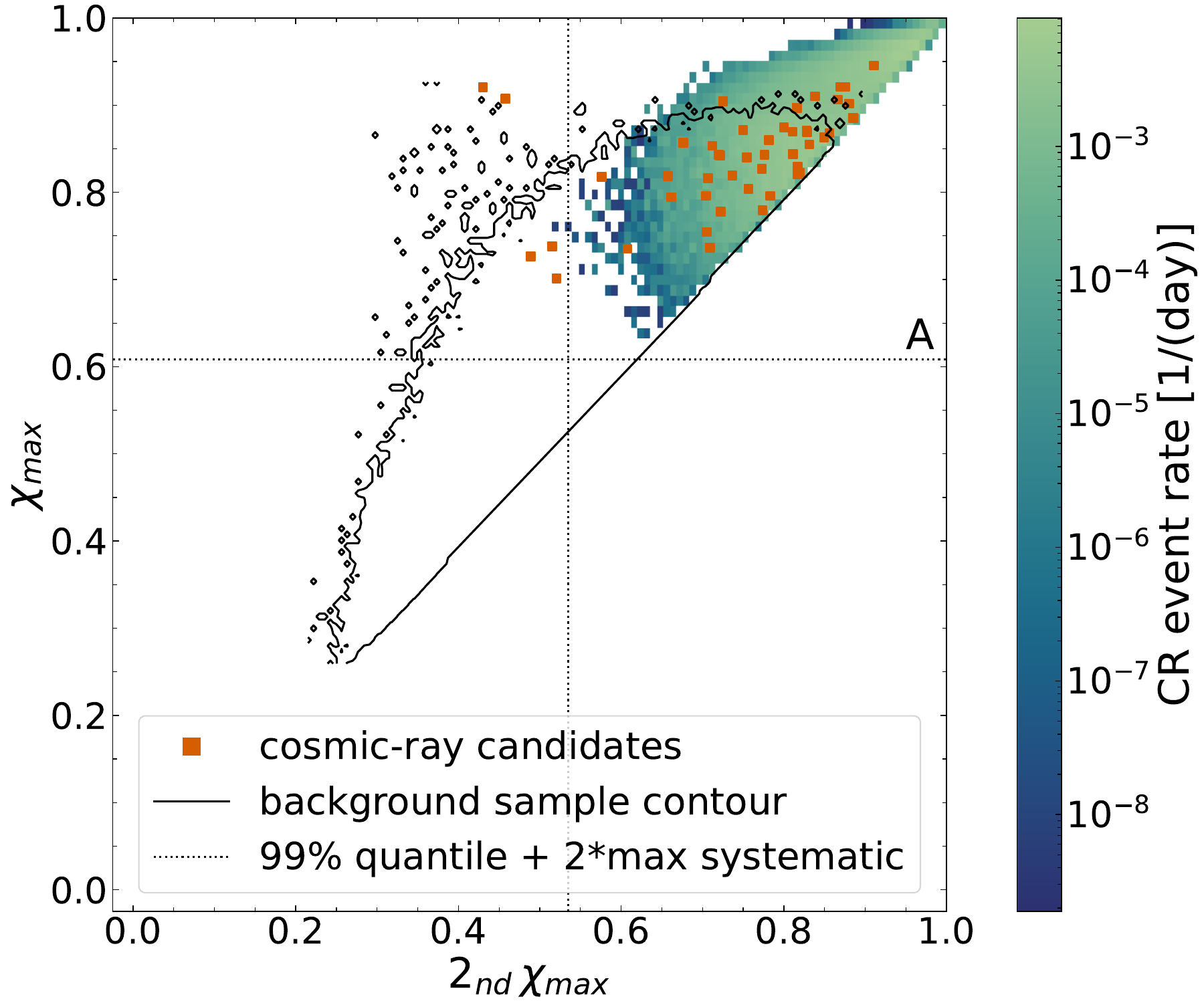}
    \caption{Correlation value of the channel with the best match as function of the correlation score of the second best channel for station 23. Shown are the contours of a background sample in black, the expected distribution of cosmic rays without systematic uncertainties (color), and the cosmic-ray candidates (squares). Region A indicates the signal region including systematics.}
    \label{fig:background}
\end{figure}

To get an additional handle on possible backgrounds, the distribution of the largest correlation score as a function of the second-largest correlation score in the same event is considered. In a few cases, background events correlate well with the template only in a single channel, which makes the second-largest correlation score a valuable quantity to consider. Also, due to the strong polarization of the UHECR signal and the triangular detector geometry, the UHECR signals are expected to not always correlate equally well for all channels, resulting in a broadened UHECR distribution.
Fig.~\ref{fig:background} shows the simulated UHECR distribution for station 23 (year 2022), in which most of the cosmic-ray candidates are found, and a conservative estimate for the systematic uncertainties. Everything within the systematic uncertainties (shown in the plot as Region A) will be considered compatible with the simulations, while background events are expected to lie outside. Additionally, the contours of the background distribution, which are all events removed by the background rejection cuts, of that station as well as the identified UHECR candidate events are displayed. The shown simulated and background distributions have a large overlap. As a result, this metric is not well-suited for signal-to-background separation, but it provides an alternative metric for cross-checks. Without a detailed treatment of systematics, we can therefore conclude that the UHECR sample of 85 candidates likely includes a minimum of 5 background events. The reconstructed parameters of these 5 events are shown separately in Fig.~\ref{fig:cosmic_ray_reconstruction}.

\subsection{Characteristics of observed cosmic rays}
\label{reconstruction_CR}
\begin{figure*}
    \centering
    \includegraphics[width=1\linewidth]{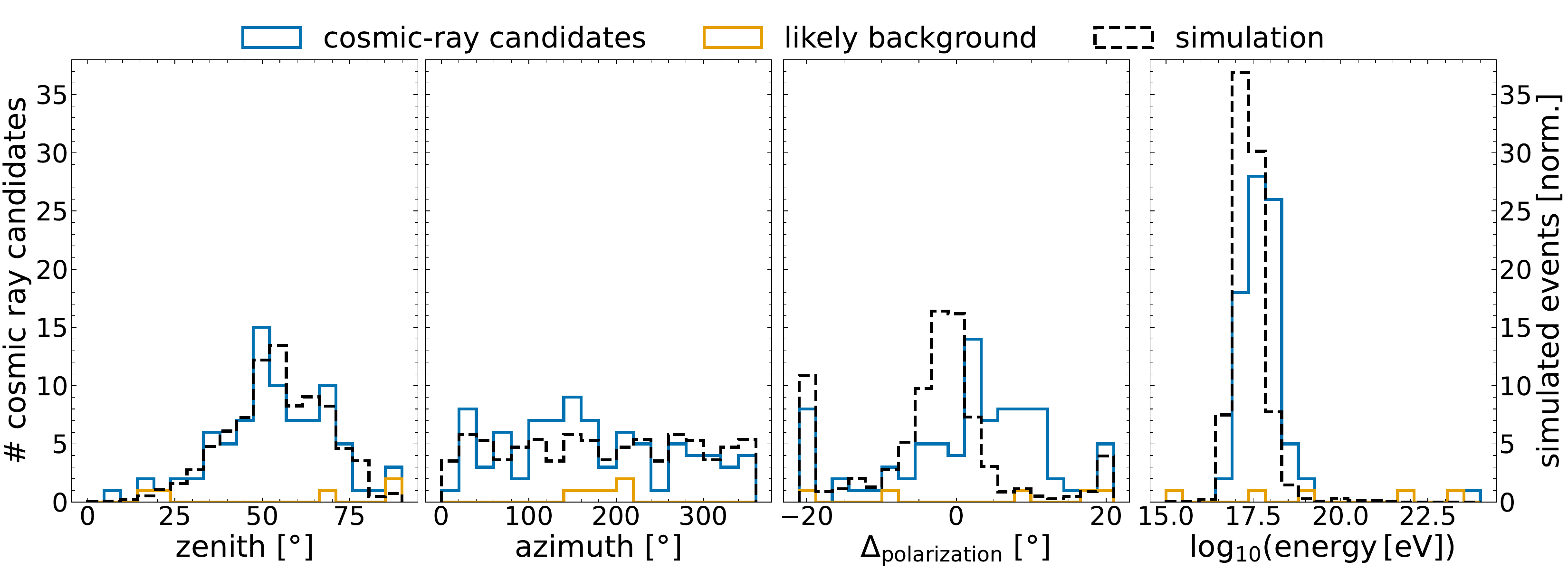}
    \caption{Reconstructed quantities for the cosmic-ray candidates (solid blue line), corresponding simulations (dashed line), which are normalized to the number of cosmic-ray candidate events, and likely backgrounds (solid orange line). The likely backgrounds are the five events identified in Sec.~\ref{sec:background}. Shown are from left to right the zenith angle, azimuth angle, difference to expected polarization angles assuming a purely geomagnetic emission (with the last bins being overflow bins), and energy. It should be noted that the shown distributions are consistent in all cases after including systematic uncertainties, which are not shown. Additionally, the simulations use on average a lower trigger threshold than realized in real data. See text for details.}
    \label{fig:cosmic_ray_reconstruction}
\end{figure*}

For each UHECR candidate event we reconstruct the zenith angle, azimuth angle, polarization and energy. The reconstruction is done using a forward folding approach, meaning a model for the electric field is fitted to the data waveforms by folding it with the hardware response, which makes it possible to estimate the zenith, azimuth and polarization together. A detailed description of this method can be found in \cite{SJOERD-THESIS}. In a second step, the energy is reconstructed using the method described in \cite{Welling:2019scz}. The resulting reconstructed parameters for all UHECR candidates can be found in Fig.~\ref{fig:cosmic_ray_reconstruction}, overlaid with a simulation that is normalized to the number of found UHECR candidate events. As discussed before, the stations all have a different trigger threshold. So we have chosen a trigger threshold for the simulation that is smaller than the thresholds of all stations to show the full possible parameter space. 

For the zenith angle, azimuth angle and energy, the reconstructed values are shown, while for the polarization, the difference ($\Delta_\mathrm{polarization}$) to the expected polarization (assuming a purely geomagnetic emission) is shown. 
The zenith and azimuth angle distributions agree well with the simulations. Due to the large distance of $1.25 \units{km}$ between two stations, only a very small number of events are expected to trigger more than one station. The number of coincidences increases with the area covered by the air-shower radio footprint, which grows with increasing zenith angle. Since some UHECR candidate events are reconstructed at a large zenith angle, we searched for potential coincidence events but found none. This result is consistent with expectations from simulations, given the small number of candidate events per station. However, we do not observe a perfect agreement for the energy. Additionally some events are reconstructed with a higher energy than expected. Due to the energy dependence of the cosmic-ray flux and the lower threshold of the simulation, more lower-energy events are included in the simulation, which - after normalizing the histogram - effectively suppresses the number of simulated higher energy events. As a result, the simulated energy distribution appears to be shifted toward slightly lower energies compared to the distribution of UHECR candidates. Furthermore, the reconstructed energies still carry sizable uncertainties, since an absolute gain calibration of the detector has not yet been performed. Much larger experiments like the Pierre Auger Observatory have measured the flux of UHECR cutting off beyond energies of $10^{20}\units{eV}$ \cite{PierreAuger:2020qqz}. Therefore, we expect that the events exceeding $10^{20}$~eV are either not true cosmic rays or simply mis-reconstructed, as no quality criteria were applied to the reconstructed dataset. Indeed, two of the highest energy events and the event below the trigger threshold are found to likely be background events (see Sec.~\ref{sec:background}).  

Also the polarization difference shows a mismatch, in the form of a bias towards positive values for the UHECR candidate events. This observed bias of $6-7^\circ$, is observed exclusively in the events from station 23, which dominate the total event sample. In contrast, the polarization differences for stations 13 and 24, although based on fewer events, agree well with the simulation. The polarization reconstruction was cross-checked by applying the method presented in Ref.~\cite{Ravn:2024arx}, which gives overall consistent results, and the same bias is present.
To estimate the impact of systematic uncertainties from the detector description and to assess whether they can explain the observed bias, we re-simulated 500 randomly selected simulated events with detector descriptions that vary slightly from the nominal detector description. These variations include geometrical shifts or rotations of the LPDA antennas that have not been accounted for during the installation of the first RNO-G stations. We then subsequently reconstruct these simulated events with the nominal (non-shifted) detector description, and determine the resulting increase in the 68\%-containment intervals and bias.

We investigated four sources of systematic uncertainty: an overall scaling of the amplifier gain, a shift of the antenna position in the x-y plane, and a rotation/tilt of the antenna orientation (either in azimuth or in zenith). For each source of systematic uncertainty, we have simulated 25 different detectors by drawing random shifts/rotations from a normal distribution, which were then used to re-simulate the selected simulated events. The shifts/rotations were applied individually to each antenna rather than globally to the entire station. The resulting biases and systematic uncertainties are summarized in Table~\ref{tab:reconstruction-systematics}. 

\begin{table}
	\centering
	\begin{tabular}{lr|rr}
		Systematic uncertainty & $\sigma$ & Bias [$^\circ$] & $\Delta \sigma_{68\%}~ [^\circ]$\\
		\hline
		Amplifier gain & 5\% & 0.0 & 0.3 \\
		Antenna position & 0.5 m & 1.3 & 2.5 \\
		Antenna azimuth & 5$^\circ$ & 3.3 & 1.9 \\
		Antenna zenith & 5$^\circ$ & 4.9 & 6.1		
	\end{tabular}
	\caption{The table shows the possible bias in reconstructed polarization and systematic uncertainty for the polarization difference (reconstructed - expected) resulting from systematic uncertainties on the detector description.}
	\label{tab:reconstruction-systematics}
\end{table}

The results in Table~\ref{tab:reconstruction-systematics} show that the observed bias can be explained by systematic uncertainties in the detector description, with the antenna azimuth and zenith angle being the biggest sources of uncertainty. Furthermore, these results emphasize that a more careful antenna position and orientation calibration for the LPDAs is needed, as it lower the systematic uncertainties of the reconstruction algorithms significantly.

As a final comparison between simulation and UHECR candidate events, we calculate the 68\% quantile of the polarization difference. For the simulation, the value strongly depends on the trigger threshold, as lower SNR events are harder to reconstruct.  If we consider a low trigger threshold for the simulation of $4\sigma$, where $\sigma$ is the RMS of the thermal noise in the trigger frequency band ($80-180 \units{MHz}$), we obtain a value of $7.3^\circ$, while for a higher trigger threshold ($14\sigma$) we obtain a value of $2.9^\circ$ \cite{SJOERD-THESIS}. To account for the bias on the polarization difference, we fit a normal distribution to the station 23 data and subtract the mean of the distribution from the data. Afterwards, the 68\% quantile is calculated to be $8.9^\circ$ for all UHECR candidate events. The values for simulation and data are compatible within the systematic uncertainties discussed above, which can also lead to a broadening of the polarization difference distribution (quoted as the difference of the 68\%-containment in Table~\ref{tab:reconstruction-systematics}). We note that a better polarization resolution has been reported by the ARIANNA collaboration also using upward-facing LPDAs \cite{ARIANNA_air_shower_polarization}. A detailed comparison of the resolutions and their limitations requires in-depth study of both detectors (different antenna configuration, trigger, waveform calibration) and differences in the methodological approach (fitting procedure, analysis cuts) and is beyond the scope of this current analysis. 

After taking into account the mentioned uncertainties, we can conclude that the reconstructed parameters of the UHECR candidates agree well with the expectation from simulations. Moreover, we want to emphasize that it is highly unlikely for a background source to reproduce all reconstructed parameter distributions. Consequently, the reconstruction supports the interpretation that most of the candidates are events from UHECR air showers. 

\subsection{Other stations}
\label{cosmic_ray_selection_other_stations}
\begin{figure*}
\begin{subfigure}{1\textwidth}
    \includegraphics[width=1\textwidth]{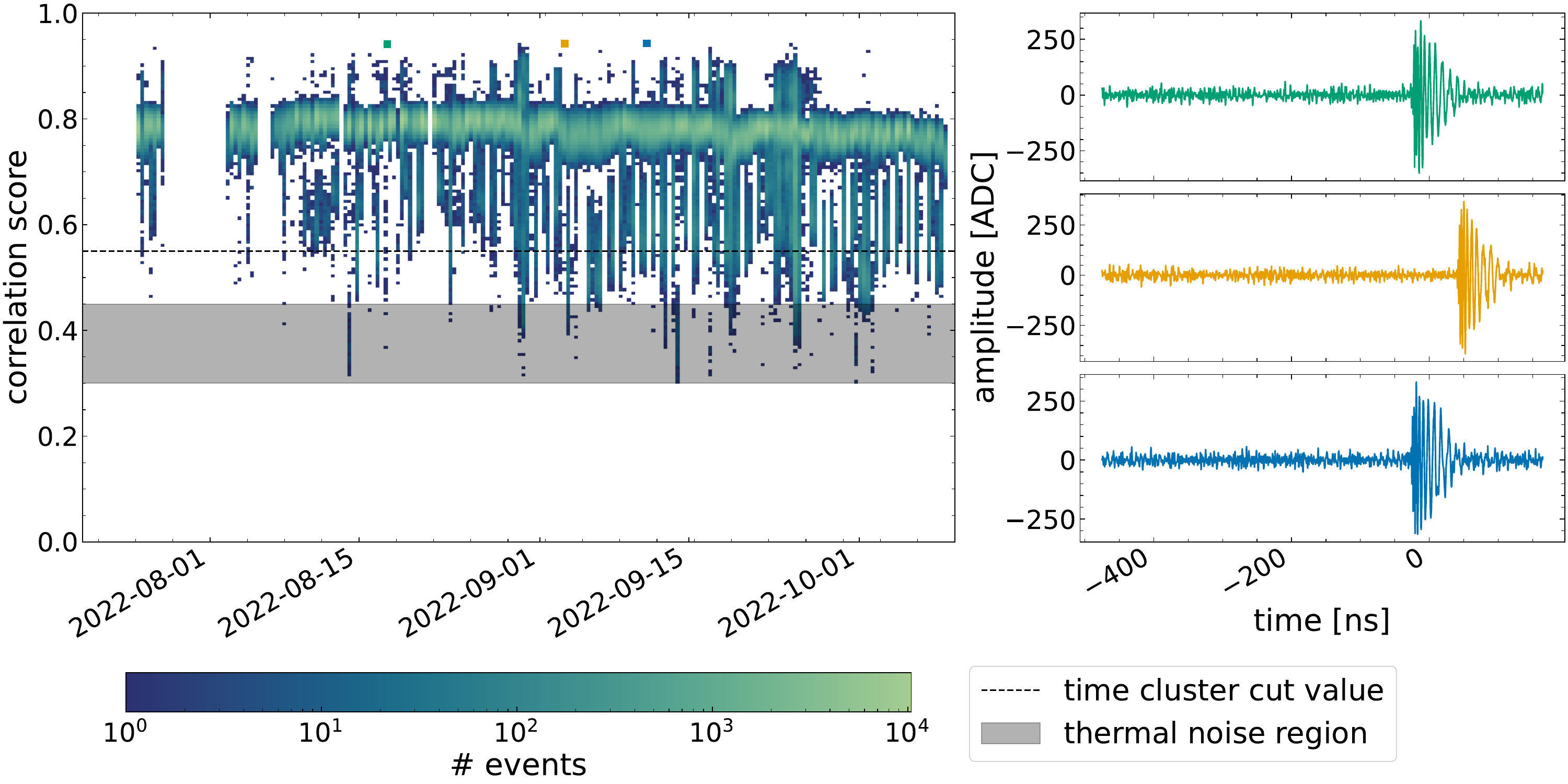}
\end{subfigure}

\begin{subfigure}{1\textwidth}
    \includegraphics[width=1\textwidth]{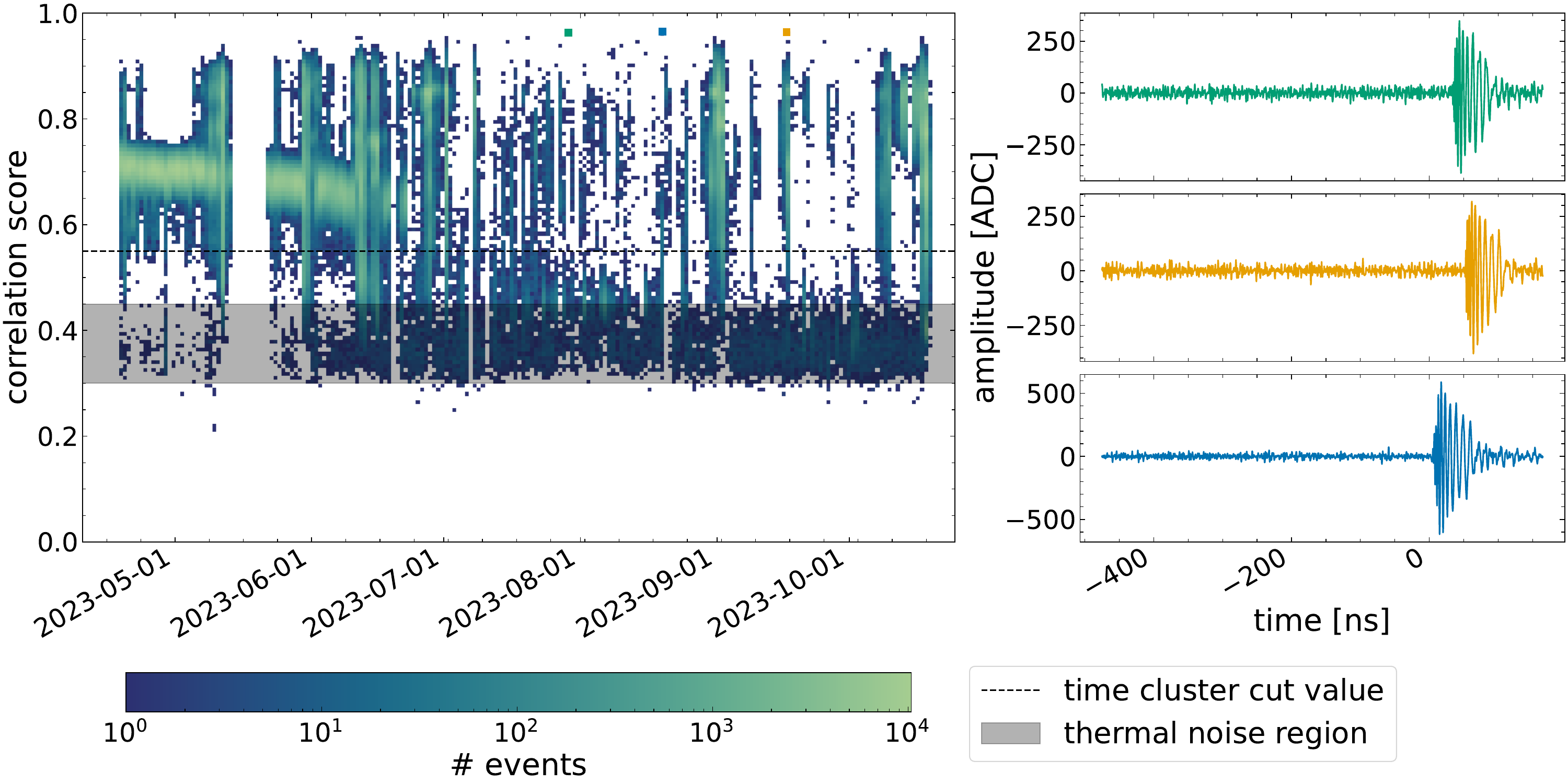}
\end{subfigure}
\caption{The figure shows the correlation score as a function of time for station 11 without any cuts applied. Additionally, it shows three example waveforms from UHECR candidate events, which are also denoted by squares and colors in the left plot. Each shown waveform corresponds to the channel with the highest correlations score per highlighted event, where the color indicates to which event it belongs. The upper plots show data from 2022, while the lower plots show 2023 data. In July 2023 changes to the power system were made.}
\label{fig:corr_vs_time_st11}
\end{figure*}

\begin{figure*}
\begin{subfigure}{1\textwidth}
    \includegraphics[width=1\textwidth]{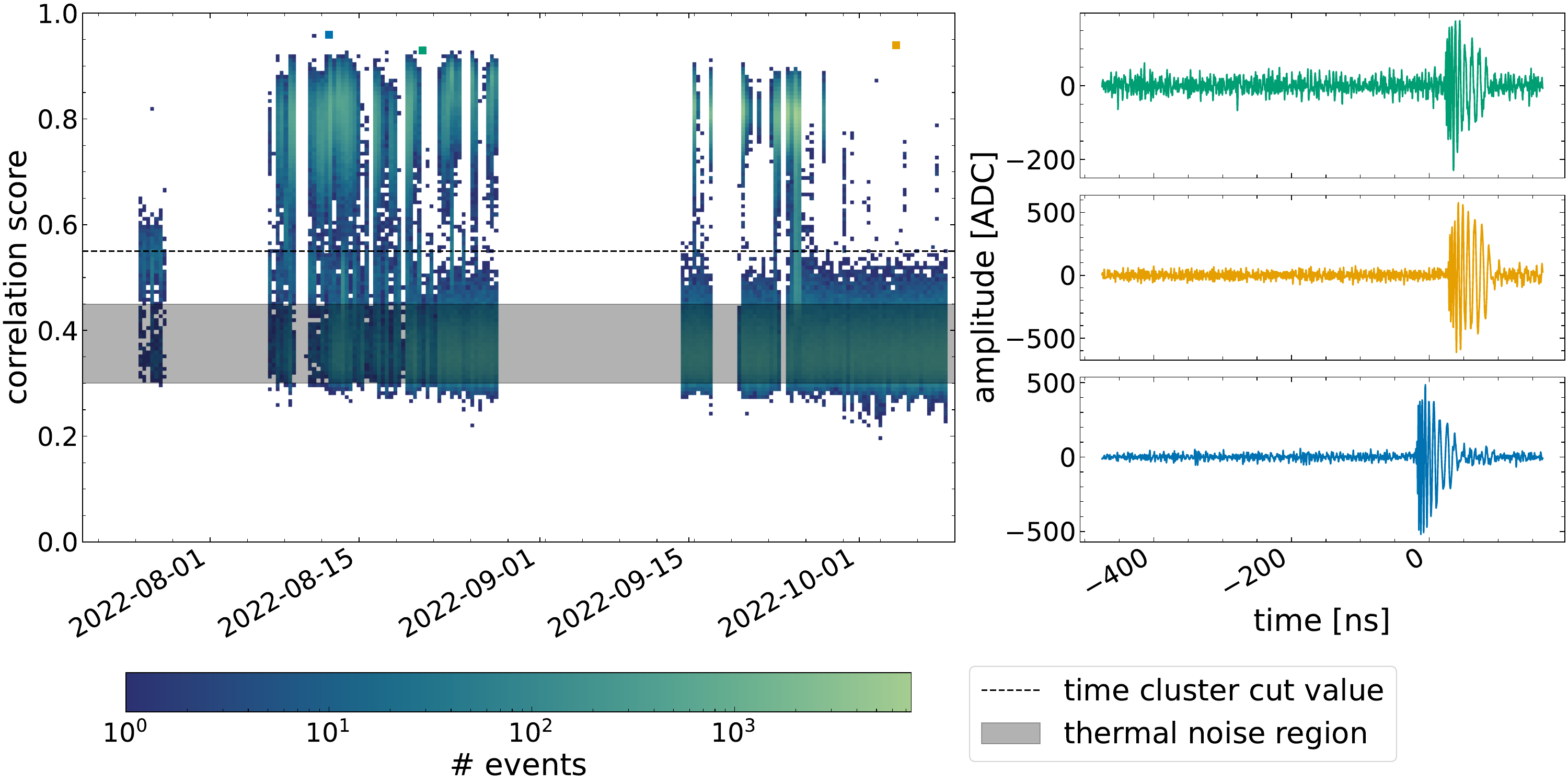}
\end{subfigure}

\begin{subfigure}{1\textwidth}
    \includegraphics[width=1\textwidth]{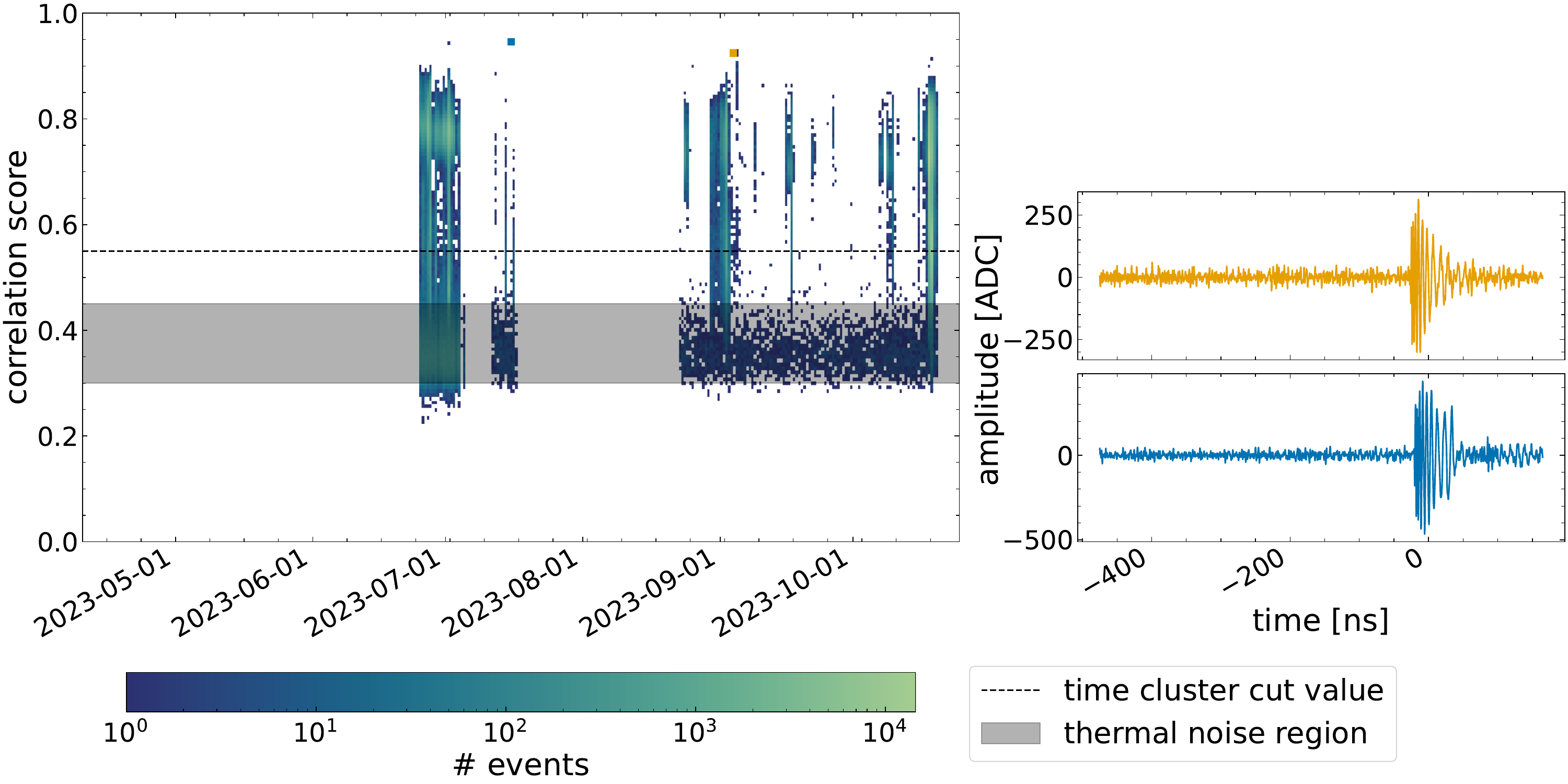}
\end{subfigure}
\caption{The same as Fig.~\ref{fig:corr_vs_time_st11}, but for station 12.}
\label{fig:corr_vs_time_st12}
\end{figure*}

\begin{figure*}
\begin{subfigure}{1\textwidth}
    \includegraphics[width=1\textwidth]{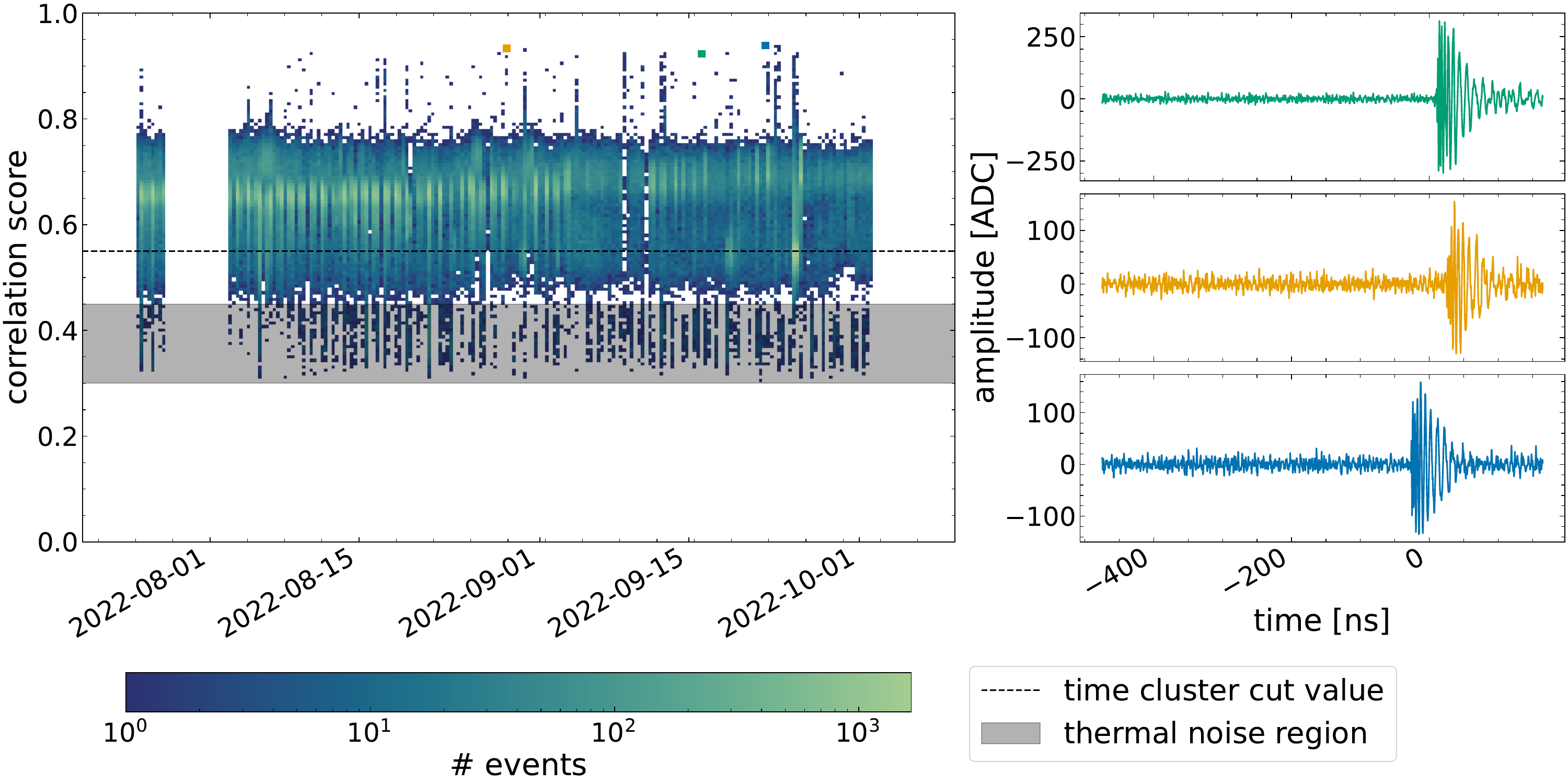}
\end{subfigure}

\begin{subfigure}{1\textwidth}
    \includegraphics[width=1\textwidth]{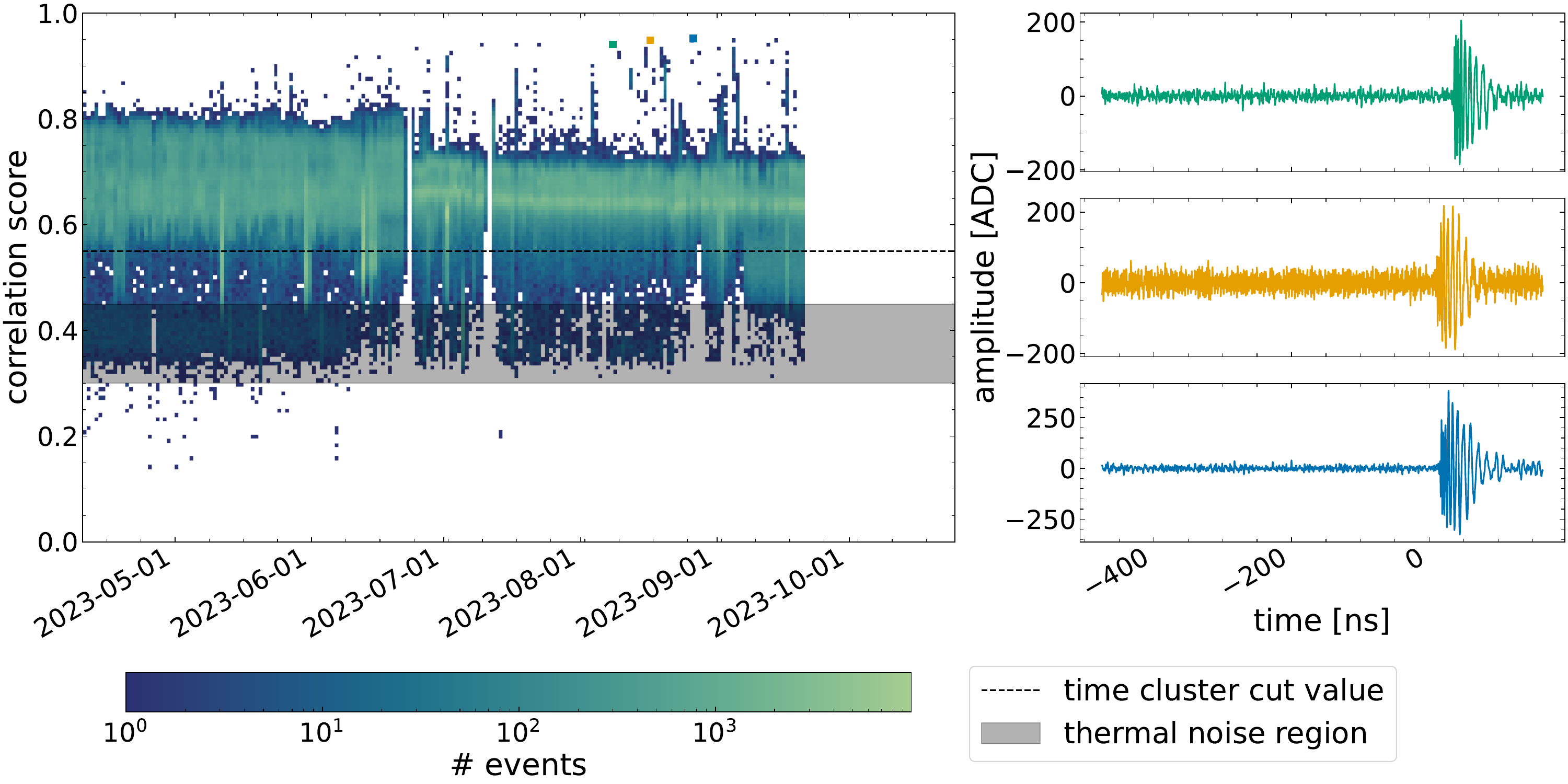}
\end{subfigure}
\caption{The same as Fig.~\ref{fig:corr_vs_time_st11}, but for station 21.}
\label{fig:corr_vs_time_st21}
\end{figure*}

\begin{figure*}
\begin{subfigure}{1\textwidth}
    \includegraphics[width=1\textwidth]{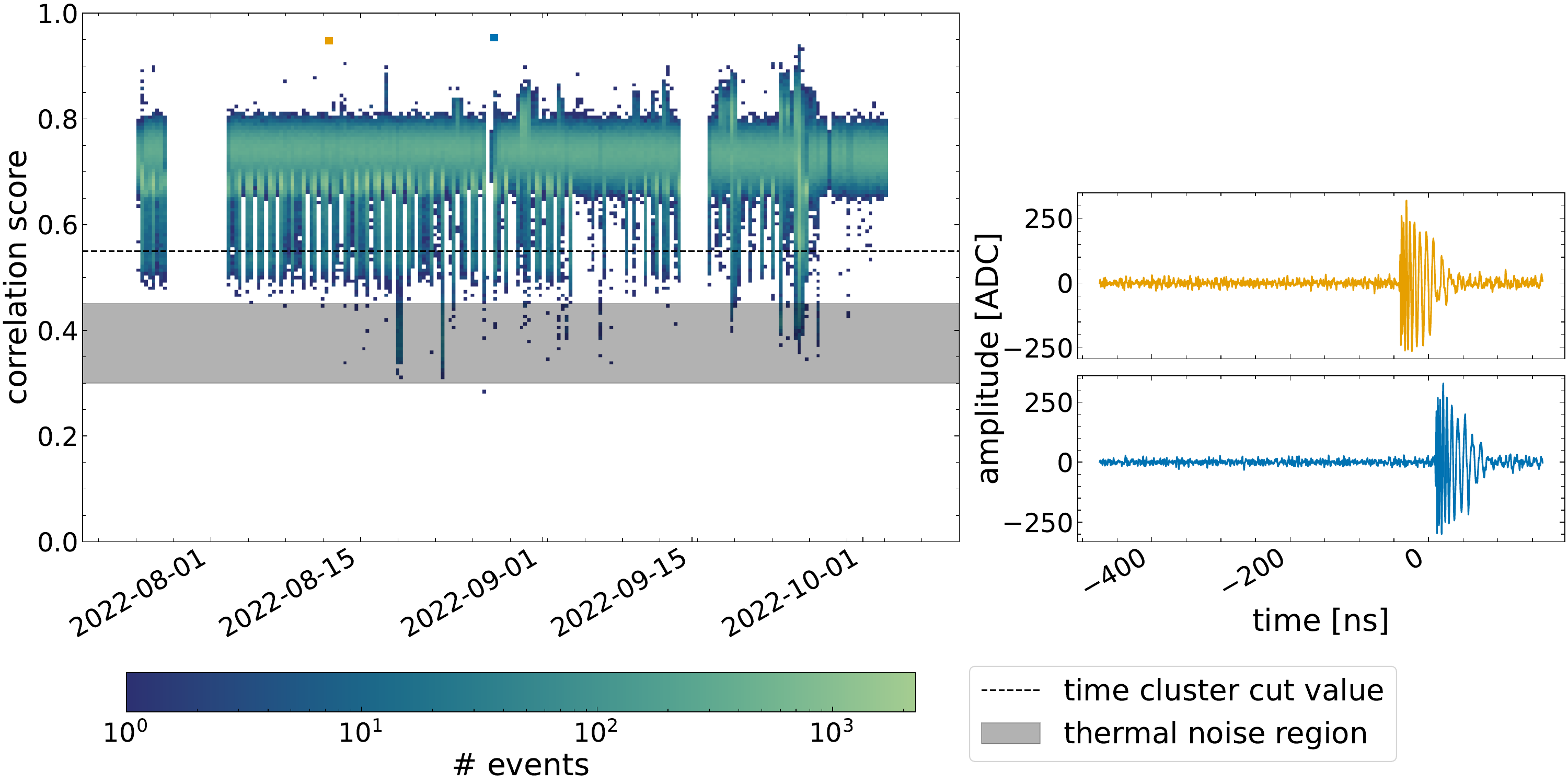}
\end{subfigure}

\begin{subfigure}{1\textwidth}
    \includegraphics[width=1\textwidth]{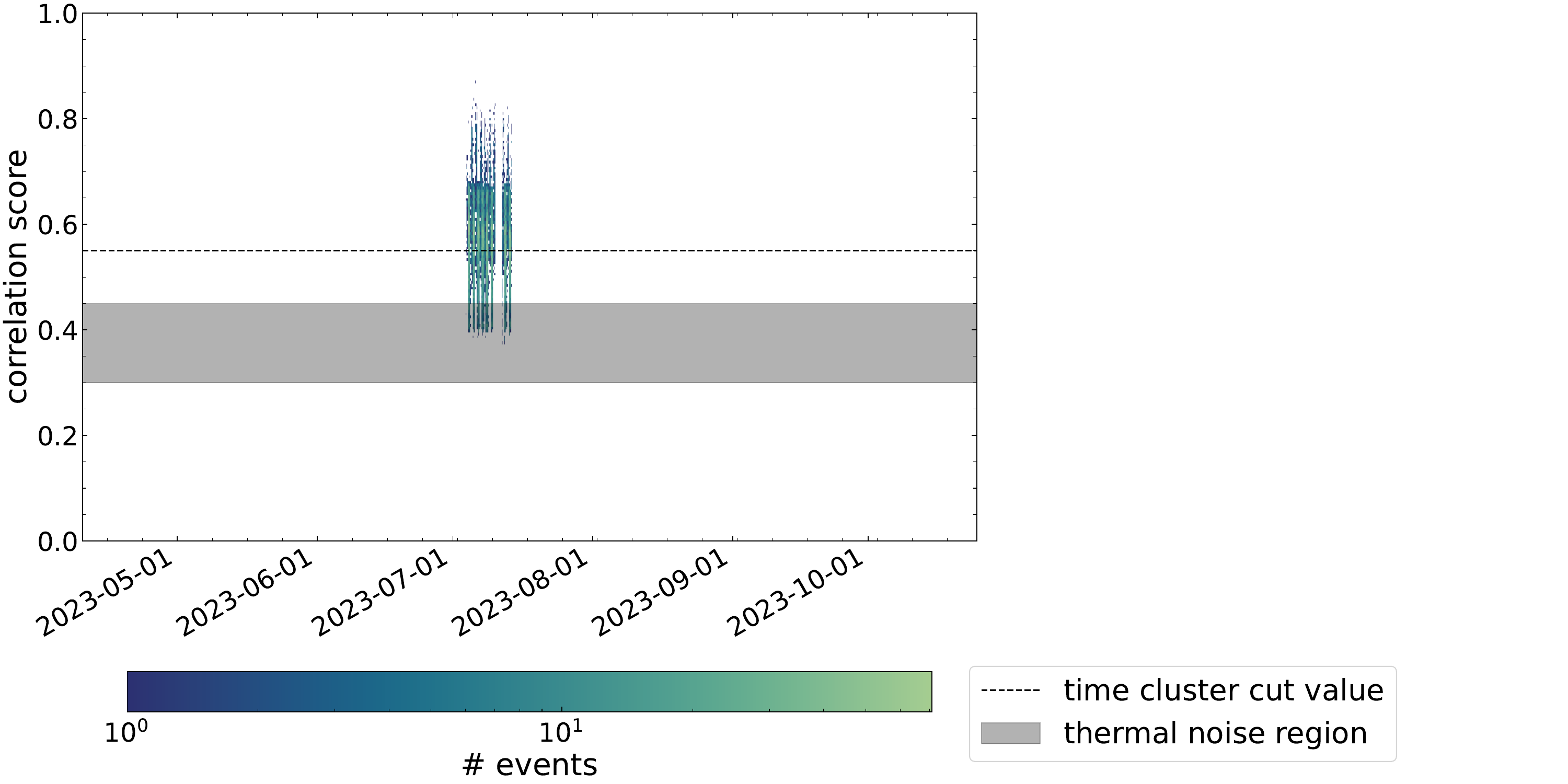}
\end{subfigure}
\caption{The same as Fig.~\ref{fig:corr_vs_time_st11}, but for station 22.}
\label{fig:corr_vs_time_st22}
\end{figure*}

As discussed in Sec.~\ref{data-set}, the stations 11, 12, 21, and 22 were excluded from the main analysis due to either an experimental wind turbine or an early version of the power system causing an unstable performance of the shallow channels. Nevertheless, we calculated the data-template correlation from all of these stations. Figures~\ref{fig:corr_vs_time_st11}, \ref{fig:corr_vs_time_st12}, \ref{fig:corr_vs_time_st21}, and \ref{fig:corr_vs_time_st22} show the correlation score as a function of time for each station and year. 


For these stations, the majority of events have a correlation score higher than the expected value for thermal noise (indicated by the grey band in the figures) and, with that, also higher than the minimal correlation score used for the time cluster cut (indicated by a black dashed line in these figures). Using the same cuts as above reduces the station's lifetime drastically. The increase of background events at higher correlation scores suggests that the noise coming from either the early test wind turbine or the early power system is impulsive, relatively similar to UHECR waveforms. This makes the shallow antennas, which are located close to these sources, challenging to use during these periods.
Upgrades in the power boxes of the two test wind turbines were implemented in early July 2023 to reduce the noise in the shallow antennas of stations 11 and 12. Figures~\ref{fig:corr_vs_time_st11} and \ref{fig:corr_vs_time_st12} show that the noise level begins to decrease around the same time, indicating that the additional shielding has a positive effect.

However, applying the time cluster cut results in lifetime losses of the order of $80\%$ for some of these stations when following the same procedure as defined in Sec.~\ref{Three_main_station_analysis}. It might be possible to design custom cuts per station to recover some of the data for a full analysis, but we decided against this approach, given that these stations are expected to have higher quality data in the coming years. In 2024, the power system was fully upgraded for the stations 21 and 22 mitigating the known issues. Additionally, the test wind turbine was replaced with a different model in 2024 and preliminary data indicates that they produce less noise in the shallow antennas. 

Although these four stations are affected by large amounts of high-correlation background, we are able to identify clear UHECR candidate events by inspecting the waveforms of the highest-correlation score events. For each station and year, the waveform of the channel with the highest correlation score is shown for up to three UHECR candidate events in Figures~\ref{fig:corr_vs_time_st11}, \ref{fig:corr_vs_time_st12}, \ref{fig:corr_vs_time_st21} and \ref{fig:corr_vs_time_st22}.
This demonstrates that, despite the issues discussed above, the stations are able to trigger on high SNR UHECR events when the up-time allows for it.

\section{Conclusion and future outlook}
In this work, we presented a search for UHECR air-showers with the shallow component of the RNO-G detector. We analyzed the full 2022-2023 dataset and reported the resulting UHECR candidate events, including their reconstructed properties. Both the number of UHECR candidates and their reconstructed properties agree with expectation from simulations. This demonstrates that, despite some necessary modifications - which were identified in this analysis - the detector performs as expected and has successfully passed the commissioning aspect using direct signals from air showers. 

We adapted the analysis method previously developed for the ARIANNA experiment by reducing the number of signal templates needed. This was achieved by utilizing the fact that the signal is primarily shaped by the antenna, amplification, and digitization response. 
In that process, we have also gained a better understanding of the impact that the amplification and digitization chain have on signal-shape-based parameters, such as the correlation score. We now have an excellent description of the amplification and digitization chain through full-chain time response measurements. For these full-chain measurements, it was found that the channel-to-channel and station-to-station fluctuations are negligible. The improved understanding and description are of immense value for future analyses that use the signal shape. 
Moreover, we thoroughly investigated various systematic effects and found that the uncertainty in the index of refraction near the surface is the dominant uncertainty when using shallow antennas. Based on different ongoing studies and new techniques (e.g. see ~\cite{probingfirnrefractiveindex}) it is reasonable to expect an improvement of this uncertainty in the near future. 

The determining factor for the energy threshold of the UHECR in the present data is the digitizer board. During the 2024 field season, the digitizer boards were upgraded for all stations, except station 24, to an improved version with a lower energy threshold. In addition, the increased noise from the power system and wind turbines was meanwhile mitigated for the stations that had to be excluded in this work. Consequently, future analyses will benefit from the upgraded station hardware and the resulting larger sample of UHECR events. We expect an increase in the number of detected cosmic rays by about an order of magnitude. 

Lastly, the newest station 14 has the same physical time delay for the shallow and deep channels, which will enable the detection of coincident signals in the shallow and deep antennas, caused by down-going cosmic rays.

\begin{acknowledgements}
We are thankful to the support staff at Summit Station for making RNO-G possible. We also acknowledge our colleagues from the British Antarctic Survey for building and operating the BigRAID drill for our project.

We would like to acknowledge our home institutions and funding agencies for supporting the RNO-G work; in particular the Belgian Funds for Scientific Research (FRS-FNRS and FWO) and the FWO programme for International Research Infrastructure (IRI), the National Science Foundation (NSF Award IDs 2411590, 2514201, 2514208, 2514206, and collaborative awards 2310122 through 2310129 and ), and the IceCube EPSCoR Initiative (Award ID 2019597), the Helmholtz Association, the German Research Foundation (DFG, Grant 556092823), the Swedish Research Council (VR, Grant 2021-05449 and 2023-00156), the University of Chicago Research Computing Center, and the European Union under the European Unions Horizon 2020 research and innovation programme (ERC, Pro-RNO-G No 101115122 and NuRadioOpt No 101116890).

\end{acknowledgements}

\bibliographystyle{JHEP}       
\bibliography{references}   


\end{document}